\documentclass[12pt]{article}

\usepackage{amssymb,amsfonts,amsthm}
\usepackage{amsmath}
\usepackage{bm}             
\usepackage[mathscr]{eucal}
\usepackage{cancel}
\usepackage{wasysym}

\def\be{\begin{equation}}
\def\ee{\end{equation}}
\def\bea{\begin{eqnarray}}
\def\eea{\end{eqnarray}}

\def\bna{\mbox{\boldmath $\nabla$}}
\def\brho{\mbox{\boldmath $\rho$}}

\def\hsp5{\hspace{5mm}}

\theoremstyle{remark}

\newcommand{\sfrac}[2]{{\textstyle{#1\over#2}}}
\newcommand{\nsum}{ \sum_{A=1}^n {}}

\newcommand{\cH}{\mathcal{H}}
\newcommand{\cC}{\mathcal{C}}
\newcommand{\cA}{\mathcal{A}}
\newcommand{\cB}{\mathcal{B}}
\newcommand{\cL}{\mathcal{L}}

\newcommand{\mv}{\mathrm{v}}
\newcommand{\mc}{\mathrm{c}}

\title{\sc  Scalar Cosmological Perturbations}

\begin{document}

\author{ \\
{\Large\sc Claes Uggla}\thanks{Electronic address:
{\tt claes.uggla@kau.se}} \\[1ex]
Department of Physics, \\
University of Karlstad, S-651 88 Karlstad, Sweden
\and \\
{\Large\sc John Wainwright}\thanks{Electronic address:
{\tt jwainwri@uwaterloo.ca}} \\[1ex]
Department of Applied Mathematics, \\
University of Waterloo,Waterloo, ON, N2L 3G1, Canada \\[2ex] }

\maketitle

\begin{abstract}

Scalar perturbations of Friedmann-Lemaitre cosmologies can be
analyzed in a variety of ways using Einstein's field equations,
the Ricci and Bianchi identities, or the conservation equations
for the stress-energy tensor, and possibly introducing a
timelike reference congruence. The common ground is the use of
gauge invariants derived from the metric tensor, the
stress-energy tensor, or from vectors associated with a
reference congruence, as basic variables.  Although there is a
complication in that there is no unique choice of gauge
invariants, we will show that this can be used to advantage.

With this in mind our first goal is to present an efficient way
of constructing dimensionless gauge invariants associated with
the tensors that are involved,  and of determining their
inter-relationships. Our second goal is to give a unified
treatment of  the various ways of writing the governing
equations in dimensionless form using gauge-invariant
variables, showing how simplicity can be achieved by a suitable
choice of variables and normalization factors. Our third goal
is to elucidate the connection between the metric-based approach and
the so-called $1+3$ gauge-invariant approach to cosmological
perturbations. We restrict our considerations to linear
perturbations, but our intent is to set the stage for the
extension to second order perturbations.

\end{abstract}

\centerline{\bigskip\noindent PACS numbers: 04.20.-q, 98.80.-k,
98.80.Bp, 98.80.Jk}

\section{Introduction}

Currently, increasingly accurate observations are driving
theoretical cosmology towards more sophisticated models of
matter and the study of possible nonlinear deviations from FL
cosmology.\footnote{We follow the nomenclature of Wainwright
and Ellis (1997):~a Friedmann-Lemaitre (FL) cosmology
 is a Robertson-Walker (RW) geometry that
satisfies Einstein's field equations.} Motivated by this state
of affairs, in a recent paper (Uggla and Wainwright (2011),
hereafter referred to as UW), we  initiated a  program of
research whose long term goal is to provide a general but
concise description of {\it nonlinear} perturbations of FL
cosmologies that will reveal the structure of the governing
equations, and hence facilitate their analysis. In furthering
this goal one is faced with making three choices. First, there
is the choice of gauge-invariant variables: the work of Bardeen
(1980) made clear that there is no unique choice. Second, the
use of dimensionless variables invariably leads to physical
insight via the choice of a suitable normalizing factor or
factors. Third, there is the choice of how to formulate the
governing equations: Einstein's field equations, the Ricci and Bianchi
identities, the conservation equation for stress-energy, or
other matter equations. In this paper we systematically
consider these three choices, working for the moment within the
framework of linear perturbation theory.

Our first goal in this paper is to present an efficient way of
constructing dimensionless gauge invariants associated with the
metric tensor, the stress-energy tensor, or other structures
that may be introduced, and of determining their
inter-relationships. We use the method of Nakamura (2003),
adapted as in UW to create dimensionless gauge invariants.

Our second goal is to give a unified treatment of  the various
ways of writing the governing equations in dimensionless form
using gauge-invariant variables within the framework of the
{\it metric-based approach} to cosmological
perturbations.\footnote{By this we mean the standard approach
to cosmological perturbations in which one  formulates the
governing equations in terms of gauge-invariant variables
associated with the perturbed metric tensor and the perturbed
stress-energy tensor, using local coordinates.} In UW we gave
the linearized Einstein equations in two forms, which we
referred to as the Poisson form, associated with the work of
Bardeen (1980), and the uniform curvature form, associated with
the work of Kodama and Sasaki (1984). In the present paper we
derive the linearized conservation equations for the
stress-energy tensor and by expressing them in terms of
suitable gauge invariants, give an alternative description of
the dynamics of scalar perturbations as a system of two first
order (in time) partial differential equations. We also include
the case where the source has multiple components. In addition,
by using the inter-relationships between the different gauge
invariants, we are able to give a unified description of the
various "conserved quantities" that are associated with long
wavelength scalar perturbations.

Our third goal is to elucidate the connection between the
metric-based approach and the so-called {\it 1+3
gauge-invariant approach} to cosmological
perturbations\footnote{The $1+3$ gauge-invariant approach uses
variables that are {\it apriori} gauge-invariant at first order
due to the Stewart-Walker lemma and are defined using the 1+3
covariant description of GR, which is based on a preferred
timelike congruence. This approach is growing in popularity.
For a recent treatment in depth we refer to Tsagas, Challinor
and Maartens (2008).}, which was developed with the goal of
circumventing the gauge difficulties associated with scalar
perturbations (Hawking (1966), Ellis and Bruni (1989) and
Ellis, Bruni and Hwang (1990)). The $1+3$ approach is
formulated independently of the metric-based
approach,\footnote{In the $1+3$ approach the metric tensor is
not used as a dynamical variable and local coordinates are not
introduced, in contrast to the metric-based approach.} and
indeed there is a significant gap between the two approaches.
In the metric-based approach it is customary to expand the
metric and other basic variables in terms of a power series in
a perturbation parameter as in UW, since this clarifies the
linearization procedure and permits one to extend the analysis
to higher order perturbations. In this respect the metric-based
approach is analogous to standard elementary perturbation
procedures in physics and engineering.  On the other hand, the
$1+3$ approach is not formulated as a conventional perturbation
procedure, and relies instead on deriving exact evolution
equations which are then linearized by dropping products of
first order terms. In this paper we will reformulate the $1+3$
approach so as to bridge the above-mentioned gap.

The plan of the paper is as follows. In section~\ref{sec:gauge}
we give the metric and stress-energy gauge invariants and
specify the four gauge choices and the two normalizations that
we will use. In section~\ref{sec:equations} we discuss the
equations for scalar perturbations that arise from the
linearization of the conservation law for the stress-energy
tensor, and their relation with the linearized Einstein
equations. The details of the derivation, which makes use of
the Replacement Principle in Appendix~\ref{app:repl},
 are given in Appendix~\ref{app:derconserved}. In
section~\ref{sec:cons} we give a concise derivation of the
so-called conserved quantities in gauge-invariant form.  In
section~\ref{sec:1+3} we introduce the basic variables in the
$1+3$ gauge-invariant approach to scalar perturbations and
derive the governing equations, which we then relate to the
corresponding equations in the metric-based approach. The details are
relegated to Appendix~\ref{app:1+3}. Section~\ref{sec:discuss}
contains a brief summary and
discussion.

\section{Gauge invariants and gauge fields}
\label{sec:gauge}

We begin by describing a dimensionless version of Nakamura's
method for constructing gauge invariants (see Nakamura (2007),
equations (2.19), (2.23) and (2.26), and UW, section 2.1, for a
brief introduction). Consider a family of tensor fields
$A(\epsilon)$ and a background scalar $\lambda$ having
dimension \emph{length} such that $\lambda^n A(\epsilon)$ is
dimensionless\footnote{See UW, footnote 9, for a discussion and
references about allocation of dimensions.}. The change induced
in the first order perturbation ${}^{(1)}\!A$ by a gauge
transformation generated by a dimensionless vector field $\xi$
on the background can be expressed using the Lie derivative
$\pounds$:
\be \label{delta_A}  \Delta {}^{(1)}\!A =
\pounds_{\xi}{}^{(0)}\!A,  \ee
(see, for example, Bruni {\it et al} (1997), equation (1.2)).
Let $X$ be a dimensionless vector field that satisfies
\be \label{delta_X}  \Delta X^a = \xi^a. \ee
It follows that the dimensionless object defined by
\be\label{bold_A}   {\bf A}[X] := \lambda^n \left({}^{(1)}\!A -
\pounds_{X} {}^{(0)}\!A \right),  \ee
is gauge-invariant. We say that ${\bf A}[X]$ is the \emph{gauge
invariant associated with ${}^{(1)}\!A$ by $X$-compensation}.
Since a choice of $X$ yields a set of gauge-invariant variables
that are associated with a specific fully fixed gauge we refer
to $X$ as the {\it gauge field}. In this paper we will use two
choices for the normalization factor $\lambda$: if $A$ is a
geometric quantity, we will use $\lambda=a$, where $a$ is the
background scale factor, while if $A$ is a matter quantity we
will use $\lambda={\cal M}$, where ${\cal M}$ is
defined\footnote{This choice is motivated in
section~\ref{sec:stress-energy}.} by~\eqref{c_M}. In the latter
case we will denote ${\bf A}[X]$ by ${\mathbb A}[X]$.

\subsection{Metric gauge invariants}

Given a 1-parameter family of metrics $g_{ab}(\epsilon)$, where
$\epsilon$ is a perturbation parameter and $g_{ab}(0)$ is a
Robertson-Walker (RW) metric, we define a dimensionless
conformal metric ${\bar g}_{ab}(\epsilon)$ according to
\be  \label{bar_g}  g_{ab}(\epsilon) = a^2 {\bar
g}_{ab}(\epsilon),   \ee
where $a$ is the scale factor of the RW metric. We expand
${\bar g}_{ab}(\epsilon)$ in powers of $\epsilon$:
\begin{equation*}  {\bar g}_{ab}(\epsilon) =
{}^{(0)} {\bar g}_{ab} + \epsilon\,  {}^{(1)} {\bar g}_{ab} + \dots\, ,
\end{equation*}
and label the unperturbed metric and (linear) metric
perturbation according to
\be\label{gamma,f}  \gamma_{ab} := {}^{(0)} {\bar g}_{ab} =
{\bar g}_{ab}(0), \qquad  f_{ab} := {}^{(1)} {\bar g}_{ab} =
\left.\frac{\partial {\bar
g}_{ab}}{\partial\epsilon}\right|_{\epsilon=0}. \ee
In order to construct a gauge field $X$ that
satisfies~\eqref{delta_X}, \emph{using only the metric}, we
need to decompose the metric perturbation $f_{ab}$ into scalar,
vector and tensor modes. Relative to a local coordinate
system\footnote{See UW equations (10) and (11).} we introduce the notation
\begin{subequations}\label{split_f}
\begin{align}
f_{00} &= -2\varphi, \\
f_{0i} &= {\bf D}_i B + B_i,\\
f_{ij} &= -2\psi \gamma_{ij} + 2{\bf D}_i {\bf D}_j C + 2{\bf D}_{(i} C_{j)} + 2C_{ij},
\end{align}
\end{subequations}
where the vectors $B_i$ and $C_i$ and the tensor $C_{ij}$
satisfy
\begin{equation*}
{\bf D}^i B_i = 0, \qquad  {\bf D}^i C_i = 0, \qquad
C^i\!_i = 0, \qquad   {\bf D}^i C_{ij} = 0,
\end{equation*}
where ${\bf D}_i$ is the spatial covariant derivative
associated with $\gamma_{ij}$. We can satisfy the \emph{spatial
part} $\Delta X^i = \xi^i$ of the requirement~\eqref{delta_X}
by choosing
\be\label{X_i}  X_i = {\bf D}_i C + C_i \ee
(UW, section 2.2), which we will take to be our default choice
for $X_i$. With this choice, the components of the gauge
invariant ${\bf f}_{ab}[X]$ associated with the metric
perturbation $f_{ab}$ by $X$-compensation, are given by
(UW, equations (21), (23) and (25))
\begin{subequations} \label{bold_f_split}
\begin{align}
{\bf f}_{00}[X] &= -2  \Phi[X] \, ,\\
{\bf f}_{0 i}[X] &= {\bf D}_i {\bf B}[X] + {\bf B}_i\, ,\\
{\bf f}_{ij}[X] &= -2\Psi[X] \gamma_{ij} + 2{\bf C}_{ij}\, .
\end{align}
\end{subequations}
where
\begin{subequations}
\be\label{metric_gi}   \Phi[X] := \varphi - (\partial_\eta +
{\cal H})X^{0},  \quad \Psi[X]  := \psi + {\cal H}X^0, \quad
{\bf B}[X] := B - \partial_\eta C + X^0 , \ee
%
%
%
%
\be\label{boldB,C} {\bf B}_i :=  B_i - \partial_\eta C_i,
\qquad   {\bf C}_{ij} := C_{ij}. \ee
\end{subequations}
In equation~\eqref{metric_gi}, ${\cal H}$ is the dimensionless
Hubble scalar, defined by\footnote{Here and elsewhere we denote
the derivative of a function $f(\eta)$ that depends only on
$\eta$ by $f'(\eta)$.  }
\be  \label{calH}  {\cal H}:= \frac{a'}{a}. \ee

The quantities ${\bf B}_i$ and ${\bf C}_{ij}$, which describe the
vector mode and tensor mode of the perturbation respectively,  are intrinsic
metric gauge invariants\footnote{Intrinsic gauge invariants
were defined in UW, section 2.1, as gauge invariants constructed solely from
a single tensor, in contrast to hybrid gauge invariants that
are constructed from several tensors.} that are independent of
the gauge field $X$. In contrast the gauge invariants $\Phi[X]$, $\Psi[X]$ and ${\bf
B}[X]$, which describe the scalar mode, depend on the choice of $X^0$
but not on the choice of the spatial gauge field $X^i$. Of course,
if we leave $X^i$ arbitrary then ${\bf f}_{ab}[X]$
 contains additional terms and its components are given by
\begin{subequations} \label{bold_f_split1}
\begin{align}
{\bf f}_{00}[X] &= -2  \Phi[X] \, ,\\
{\bf f}_{0 i}[X] &= {\bf D}_i {\bf B}[X] + {\bf B}_i + \partial_{\eta}{\bf Z}_i[X]\, ,\\
{\bf f}_{ij}[X] &= -2\Psi[X] \gamma_{ij} + 2{\bf C}_{ij} + 2{\bf D}_{(i} {\bf Z}_{j)}[X]\, .
\end{align}
\end{subequations}
where
\be {\bf Z}_i[X]:={\bf D}_i C + C_i - X_i. \ee
Our default choice~\eqref{X_i} for $X^i$ corresponds to ${\bf Z}_i[X]=0$.

\subsection{Stress-energy gauge invariants}
\label{sec:stress-energy}

Consider a stress-energy tensor $T^a\!_b(\epsilon)$ that obeys
the background symmetries, {\it i.e.}, it is spatially homogeneous
and isotropic:
\begin{equation}\label{T_0} {\bf D}_i{}^{(0)}\!T^\alpha\!_\beta = 0, \qquad
{}^{(0)}\!T^0\!_i = {}^{(0)}\!T^i\!_0 = 0 ,\qquad
{}^{(0)}\!T^i\!_j = \sfrac13\,\delta^i\!_j\,{}^{(0)}\!T^k\!_k .
\end{equation}
We assume that $T^a\!_b(\epsilon)$ satisfies the conservation
law $\,{}\!^\epsilon\bna\!_b T^b\!_a(\epsilon) = 0$, which at
zeroth order yields
\begin{subequations} \label{conserved0}
\begin{equation}
{}^{(0)}\!\rho^\prime = -3{\cal H}\left({}^{(0)}\!\rho\, + {}^{(0)}\!p\right) ,
\end{equation}
where
\begin{equation}
{}^{(0)}\!\rho = -{}^{(0)}\!T^0\!_0,\qquad {}^{(0)}\!p = \sfrac13 {}^{(0)}\!T^k\!_k .
\end{equation}
\end{subequations}

When constructing dimensionless gauge invariants it is
necessary to choose a normalization factor. In Newtonian theory
the dimensionless quantity $\delta\rho/\rho\equiv{}^{(1)}\!\rho/{}^{(0)}\!\rho$,
where $\rho$ is the mass density, is
 used to describe structure formation.  By
analogy the same quantity is usually used in GR, but with
$\rho$ being the mass-energy density instead. We propose that
in GR a more natural normalization factor is the inertial
mass-energy density ${}^{(0)}\!\rho + {}^{(0)}\!p$, since this
is the quantity that appears instead of $^{(0)}\!\rho$ in the
relativistic energy-momentum conservation equations.
%
%
As we will show in this paper normalizing with ${}^{(0)}\!\rho
+ {}^{(0)}\!p$ leads to a simpler description of scalar density
perturbations when using matter variables. We shall refer to
this type of normalization as \emph{inertial mass-density
normalization} or more briefly, as {\it ${\cal
M}$-normalization}.

In order to implement the above idea we assume that the
inertial mass-density ${}^{(0)}\!\rho\, + {}^{(0)}\!p$
in~\eqref{conserved0} is positive, and introduce a
normalization factor with dimension $length$, defined by
\begin{equation}\label{c_M}
{\cal M} := \left({}^{(0)}\!\rho\, + {}^{(0)}\!p\right)^{-1/2} .
\end{equation}
As in UW we introduce the notation
\be  \label{A_T}  \cA_T := a^2({}^{(0)}\!\rho + {}^{(0)}\!p),
\qquad {\cC_T^2} :=\frac{{}^{(0)}\!p'}{{}^{(0)}\!\rho'}, \ee
and in analogy with UW\footnote{Replace $A$ by $T$ in equations
(39) and (40) in UW.} we define the following intrinsic gauge
invariants associated with the stress-energy
tensor\footnote{The $a$-normalized gauge invariants ${\bf
T}^a\!_b$ in UW are related to the corresponding ${\cal
M}$-normalized gauge invariants $\mathbb{T}^a\!_b$ via ${\bf
T}^a\!_b = {\cal A}_T\mathbb{T}^a\!_b$, where ${\cal A}_T =
a^2({}^{(0)}\!\rho + {}^{(0)}\!p) = (a/{\cal M})^2$.}, using
$\cal M$-normalization:
\begin{subequations}\label{GT}
\begin{align}
\hat{\mathbb T}^i\!_j &:=  {\cal M}^2\,{}^{(1)}\!{\hat T}^i\!_j \\
{\mathbb T}_i &:= -  {\cal M}^2\left( {\bf D}_i {}^{(1)}\!T^0\!_0 +
3{\cal H}{}^{(1)}\!T^0\!_i \right), \label{GiTi}\\
{\mathbb T} &:= {\cal M}^2({\cal C}_T^2
{}^{(1)}\!T^0\!_0 + \sfrac13 {}^{(1)}\!T^k\!_k)  \label{bf_T1},
\end{align}
\end{subequations}
where
\be  \label{GT_hat}  {}^{(1)}\!{\hat T}^i\!_j:= {}^{(1)}\!T^i\!_j -
\sfrac13\delta^i\!_j {}^{(1)}\!T^k\!_k.  \ee
We also introduce the following
 gauge invariants by $X$-compensation\footnote
{Replace $A$ by $T$ in equations (38a) and (38b) in UW
and multiply by $1/{\cal A}_T = \left({\cal M}/a\right)^2$.
These equations also arise from~\eqref{bold_A} with
$A$ replaced by $T$ and $\lambda = {\cal M}$.}
using $\cal M$-normalization:
\begin{subequations}  \label{TXcomp}
\begin{align}
\mathbb{T}^0\!_0[X] &:= {\cal M}^2\left({}^{(1)}\!T^0\!_0\right) -
3\cH X^0 \label{DX}\\
\mathbb{T}^0\!_i[X] & := {\cal M}^2\,{}^{(1)}\!T^0\!_i
+ {\bf D}_i X^0. \label{TiX}
\end{align}
\end{subequations}
It follows from~\eqref{GiTi} and~\eqref{TXcomp} that
\be \label{BbbT_i}   {\mathbb T}_i = - \left( {\bf D}_i {\mathbb T}^0\!_0[X] +
3{\cal H} {\mathbb T}^0\!_i[X] \right).  \ee
Note that stress-energy gauge invariants depend only on the
choice of $X^0$, not on $X^i$.

In analogy with UW (see equations (50)) we  decompose the
matter gauge invariants ${\hat{\mathbb T}}{}^i\!_j, {\mathbb T}_i, {\mathbb
T}, {\mathbb T}^0\!_i[X]$ and ${\mathbb T}^0\!_0$ into scalar, vector, and tensor
modes and label them as follows:
\begin{subequations}  \label{T_i}
\begin{align}
{\hat{\mathbb T}}{}^i\!_j &= {\bf D}^i\!_j{\bar\Pi} +\,
2\gamma^{ik}{\bf D}_{(k}{\bar\Pi}_{j)} + {\bar{\Pi}}^i\!_j , \label{bf_T_hat}\\
{\mathbb T}_i & = {\bf D}_i {\mathbb D}+ {\mathbb D}_i, \label{bf_T_i} \\
{\mathbb T} & = {\bar \Gamma}, \label{bf_T}\\
{\mathbb T}^0\!_i[X] & = {\bf D}_i {\mathbb V}[X] + {\mathbb V}_i, \label{hybrid_T}  \\
{\mathbb T}^0\!_0[X] &= -{\mathbb D}[X], \label{T_00}
\end{align}
where
\begin{equation}  \label{restrict}
{\bf D}^i{\bar \Pi}_i =0 ,\qquad {\bar {\Pi}}^k\!_k = 0 ,\qquad  {\bf
D}_i{\bar {\Pi}}^i\!_j = 0, \qquad {\bf D}^i {\mathbb D}_i = 0, \qquad
{\bf D}^i {\mathbb V}_i=0,
\end{equation}
and
\be\label{Dij}    {\bf D}_{ij} := {\bf D}_{(i}{\bf D}_{j)} -
\sfrac13\gamma_{ij}{\bf D}^2, \qquad {\bf D}^2 := {\bf D}^i{\bf
D}_i. \ee
\end{subequations}
It follows from~\eqref{BbbT_i} and~\eqref{T_i} that
\begin{equation}
\mathbb{D} = \mathbb{D}[X] - 3\cH\mathbb{V}[X],
\qquad \mathbb{D}_i = -3\cH\mathbb{V}_i .
\end{equation}

\subsection{Standard choices of the gauge field}
\label{subsection:choice}

In order to eliminate the gauge freedom in the scalar mode,
thereby determining the perturbed metric uniquely, we have to fully
specify the gauge field $X$. We fix the spatial part
$X^i $ of the gauge field {\it ab initio} as in
equation~\eqref{X_i}, leaving the temporal part $X^0$ to be
specified. We observe that $X^0$ appears {\it linearly and
algebraically} in the definitions of the gauge invariants:
\be \label{basic_gi} \Psi[X], \quad {\bf B}[X], \quad {\mathbb D}[X],  \quad  {\mathbb V}[X].  \ee
Note that $\Psi[X]$ and ${\bf B}[X]$ are defined by~\eqref{metric_gi},
while ${\mathbb D}[X]$ and ${\mathbb V}[X]$ are given by\footnote
{In deriving the expression for ${\mathbb V}[X]$, we assume, as in UW,
that the inverse operator of ${\bf D}^2$ exists.
In terms of the $(1+3)$-decomposition of the stress-energy tensor,
we can write ${\mathbb V}[X] = v+{\bar{\mathbb Q}} + X^0$, as follows from
\eqref{TvQ} and~\eqref{vQ_decomp}.}
\be \label{bbDV} {\mathbb D}[X]= - {\cal M}^2 {^{(1)}}T^0\!_0
+3{\cal H}X^0, \qquad {\mathbb V}[X] = {\cal M}^2{\bf D}^{-2}
{\bf D}^i {^{(1)}}T^0\!_i + X^0, \ee
as follows from~\eqref{TXcomp},~\eqref{hybrid_T},~\eqref{T_00}
and~\eqref{restrict}. We can thus determine $X^0$ {\it
uniquely} by requiring that one of these four variables be
zero. These choices in fact correspond to four of the commonly
used gauges in cosmological perturbation theory.\footnote{These
gauge choices and others are discussed, for example, by Kodama
and Sasaki (1984), Hwang (1991), Hwang and Noh (1999) and Malik
and Wands (2009). In contrast to our approach which emphasizes
relations between gauge invariants, they define gauge
invariants in terms of gauge-variant quantities.}
\begin{itemize}
\begin{subequations}\label{gaugechoices}
\item[(i)] Poisson gauge:
\begin{equation}
\qquad {\bf B}[X_\mathrm{p}] = 0.
\end{equation}
\item[(ii)] Uniform curvature gauge:
\begin{equation}  \label{X_c}
\qquad \Psi[X_\mathrm{c}] = 0.
\end{equation}
\item[(iii)] Total matter gauge:
\begin{equation}  \label{X_V}
\qquad\,\, \mathbb{V}[X_\mathrm{v}] = 0.
\end{equation}
\item[(iv)] Uniform density gauge:
\begin{equation}  \label{X_rho}
\qquad\,\,\, \mathbb{D}[X_\rho] = 0.
\end{equation}
\end{subequations}
\end{itemize}

Determining $X^0$ in this way does in fact satisfy condition
\eqref{delta_X}, $\Delta X^0=\xi^0$. This has been verified
for $X_{\mathrm p}$ and $X_{\mathrm c}$ in UW (see equation (26)).
For the other two cases, we need the transformation laws:
\be \label{trans} \Delta ^{(1)}T^0\!_0 = -3{\cal H}{\cal M}^{-2} \xi^0, \qquad
  \Delta ^{(1)}T^0\!_i = -{\cal M}^{-2} {\bf D}_i \xi^0,  \ee
  which are a consequence of~\eqref{delta_A}.\footnote
{Note that the formal similarity between~\eqref{delta_A} and~\eqref{bold_A}
enables one to obtain~\eqref{trans} directly from~\eqref{TXcomp} without any
calculation.}  Condition~\eqref{delta_X} now follows immediately
from \eqref{bbDV},~\eqref{X_V},~\eqref{X_rho} and~\eqref{trans}.

In practice, we will not use the explicit expressions for $X^0$ that are
defined implicitly by equations~\eqref{gaugechoices}. Instead,
in order to be able to relate gauge invariants associated with
different choices of gauge field we introduce a set of {\it transition
rules}. Let $X^0_{\bullet}$ be a specific choice of the temporal
gauge field and let $X^0$ be an arbitrary choice. The
difference
\begin{equation}  \label{Zbullet1}
{\bf Z}^0_\bullet[X] := X^0_\bullet - X^0,
\end{equation}
is gauge-invariant on account of~\eqref{delta_X}. The
desired transition rules are as follows:
\begin{subequations}  \label{transition}
\begin{xalignat}{2}
\Psi[X] &= \Psi[X_\bullet]  - {\cal H}{\bf Z}^0_\bullet[X], &\,
{\bf B}[X] &= {\bf B}[X_\bullet] - {\bf Z}^0_\bullet[X], \label{metricinv}\\
\mathbb{D}[X] &= \mathbb{D}[X_\bullet] - 3\cH {\bf
Z}^0_\bullet[X], &\, \mathbb{V}[X] &= \mathbb{V}[X_\bullet] -
{\bf Z}^0_\bullet[X]. \label{stressenergyinv}
\end{xalignat}
\end{subequations}
Equations~\eqref{metricinv} follow immediately
from~\eqref{metric_gi}, while equations~\eqref{stressenergyinv}
are a consequence of~\eqref{TXcomp},~\eqref{hybrid_T}
and~\eqref{T_00}.

By inspection of~\eqref{transition} we see that the following
linear combinations of the variables~\eqref{basic_gi} are {\it
independent of the choice of $X$}:
\begin{subequations}\label{indep_X}
\begin{xalignat}{2}
[\Psi, {\bf B}] &:= \Psi[X] -\cH {\bf B}[X], &\,
[\Psi,{\mathbb V}] &:= \Psi[X] -\cH {\mathbb V}[X],\\
[{\mathbb V},{\bf B}] &:=  {\mathbb V}[X] - {\bf B}[X], &\,
[{\mathbb D}, {\mathbb V}] &:= {\mathbb D}[X] - 3\cH {\mathbb V}[X],\\
[{\mathbb D}, {\bf B}] &:= {\mathbb D}[X] - 3\cH {\bf B}[X],
&\, [{\mathbb D},\Psi] &:= {\mathbb D}[X] - 3\Psi[X].
\end{xalignat}
\end{subequations}
We can thus substitute two different choices of $X^0$ into any
of the $X$-independent expressions in~\eqref{indep_X} and equate
the results, thereby relating different gauge invariants. For
example, if we first choose $X^0 = X^0_{\mathrm c}$ in $[{\mathbb
D},\Psi]$ and then keep $X^0$ arbitrary as the second choice,
we obtain
\be {\mathbb D}[X_{\mc}] = {\mathbb D}[X] - 3 \Psi[X],  \ee
on account of~\eqref{X_c}. If we set $X^0 = X^0_{\rho}$
in this equation it follows that
\be \label{DcPsirho} {\mathbb D}[X_{\mc}] = - 3 \Psi[X_{\rho}],  \ee
on account of~\eqref{X_rho}.

The gauge invariant $\Phi[X]$ is on a different footing from
the gauge invariants~\eqref{basic_gi} since it depends on the
derivative of $X^0$ through equation~\eqref{metric_gi}. Thus
requiring $\Phi[X]=0$ does not determine $X^0$ uniquely and
hence does not lead to a fully fixed gauge\footnote{As a
consequence the synchronous gauge contains residual freedom.}.
Nevertheless, $\Phi[X]$ does have a well-defined transition
rule analogous to~\eqref{transition}, namely
\be \label{transition2} \Phi[X] = \Phi[X_{\bullet}] +
(\partial_\eta + \cH){\bf Z}^0_{\bullet}[X],  \ee
as follows from~\eqref{metric_gi}. By
comparing~\eqref{transition2} with~\eqref{transition} one can
construct $X$-independent linear combinations of $\Phi[X]$ and
the variables in~\eqref{basic_gi}, analogous
to~\eqref{indep_X}. For example,
\be \label{indep_X2} [\Phi,{\mathbb V}]:= \Phi[X] +  (\partial_\eta + \cH){\mathbb V}[X], \ee
is independent of $X$. There are three other expressions
linking $\Phi[X]$ with $\Psi[X], {\bf B}[X]$ and ${\mathbb D}[X]$
that can be written if needed. If we set $X=X_{\mathrm v}$
in~\eqref{indep_X2} and use~\eqref{X_V} we obtain
\be \label{Phi_V}  \Phi[X_{\mathrm v}] = \Phi[X] +  (\partial_\eta + \cH){\mathbb V}[X], \ee
which we will use later.

To conclude this section we note that the gauge invariants and
$X$-independent combinations that we have introduced do not
exhaust all possibilities, but do serve to illustrate an
efficient way of defining gauge invariants and determining
their inter- relationships, which constitutes one of the main
results of this paper. A further example arises in
appendix~\ref{app:1+3metric}, where we make use of another
gauge invariant, namely the linear perturbation of the Hubble
scalar of a timelike reference congruence, denoted by ${\bf
H}[X]$. In working with this gauge invariant we find it
necessary to introduce an $X$-independent combination involving
three gauge invariants (see equation~\eqref{PsiPhiV}).

\subsubsection*{Notation}

Our general notation for dimensionless gauge invariants is
exemplified by $\Psi[X_\mathrm{c}]$ and $\mathbb{V}[X_{\mathrm
p}]$, {\it i.e.} a capital letter, or a bold face letter, or a
special font, e.g. $\mathbb{V}$, which denotes inertial
mass-density normalization, replaces the symbol for an
associated gauge-variant variable, with the choice of the gauge
vector field indicated by a subscript on the symbol $X$. For
convenience we will often simplify the notation by setting
$\Psi[X_\bullet]=\Psi_\bullet$, {\it etc.}. For some of the
commonly used gauge invariants we will use unsubscripted
symbols:
\begin{subequations} \label{gi_notation}
\begin{xalignat}{3}
\Phi &:= \Phi[X_\mathrm{p}],&\, \Psi &:= \Psi[X_\mathrm{p}],&\,
\mathbb{V} &:=
\mathbb{V}[X_\mathrm{p}],\\
{\bf A} &:= \Phi[X_\mathrm{c}],&\, {\bf B} &:= {\bf
B}[X_\mathrm{c}],&\, \mathbb{D} &:= \mathbb{D}[X_\mathrm{v}].
\label{DV}
\end{xalignat}
\end{subequations}
%

\section{ Structure of the linearized governing equations}
\label{sec:equations}

In this section we give different forms for the governing
equations for scalar perturbations, first using the metric
gauge invariants as basic variables, and then using the
stress-energy gauge invariants.

\subsection{Linearized Einstein field equations}
\label{sec:lin_einst}

As shown in UW there are two natural choices of intrinsic \emph{metric} gauge
invariants when formulating the linearized Einstein equations
for scalar perturbations, the {\it uniform curvature} gauge
invariants and the {\it Poisson} gauge invariants. As in UW we
introduce the geometric background scalars ${\cal A}_G$ and
${\cal C}_G^2$, with ${\cal C}_G^2$ defined in terms of the
derivative of ${\cal A}_G$:
\be {\cA}_G := 2(-\cH' + \cH^2 + K), \quad
 \cA_G' = -(1 + 3\cC_G^2)\cH\cA_G ,  \label{A_G}  \ee
(see UW, equation (42)). With ${\cal A}_T$ and
${\cal C}_T^2$ defined by~\eqref{A_T}, the background
Einstein equations imply that $\cA_G = \cA_T$
and ${\cal C}_G^2 = {\cal C}_T^2$. We denote their common values
by $\cA$ and ${\cal C}^2$:
\be\label{cal_C} \cA = \cA_G = \cA_T, \qquad  {\cal C}^2 =
{\cal C}_G^2 = {\cal C}_T^2. \ee

\subsubsection*{The uniform curvature formulation}

The governing equations in the uniform curvature formulation
are\footnote{See UW, equation (52). We give the matter terms on
the right hand side in two forms: using $a$-normalization as in
UW, and ${\cal M}$-normalization as introduced in the present
paper.}:
\begin{subequations} \label{scalar_eq_curv}
\begin{alignat}{2}
\cL_B{\bf B} + {\bf A} &=&\,\, - \Pi\qquad &= - \cA_T\bar{\Pi}\,\label{bfB_evol}\\
\cH\!\left(\cL_A{\bf A} + {\cal C}_G^2 {\bf D}^2{\bf B}\right)
&=&\,\, \sfrac12\Gamma + \sfrac13{\bf D}^2\Pi &=
\cA_T(\sfrac12\bar{\Gamma} + \sfrac13{\bf D}^2\bar{\Pi}), \label{Phicurv_evol}\\
{\cal H}\!\left({\bf D}^2 + 3K\right){\bf B} &=&\,\, -
\sfrac12\Delta \qquad &= - \sfrac12\cA_T\mathbb{D}, \label{Poisson_C} \\
\cH {\bf A} + (\sfrac12\cA_G - K){\bf B} &=&\,\, - \sfrac12 V \qquad &=
- \sfrac12\cA_T\mathbb{V}, \label{V_F}
\end{alignat}
\end{subequations}
where the first order differential operators ${\bf{\cal L}}_A$
and  ${\bf{\cal L}}_B$ are defined by
\begin{subequations}
\be \cL_A := \partial_\eta + \cH\cB, \qquad \cL_B :=
\partial_\eta + 2\cH,  \label{L_B} \ee
with
\be  \cB := \frac{2{\cH}'}{ {\cal H}^2} + 1 + 3\cC_G^2.  \ee
For future reference we note that
\be \label{cB} \cH\cB =
-\left(\frac{\cA_G}{\cH^2}\right)^{-1}\left(\frac{\cA_G}{\cH^2}\right)^\prime,  \ee
\end{subequations}
as follows from~\eqref{A_G}.

\subsubsection*{The Poisson formulation}

The governing equations in the Poisson formulation are (UW,
equations (54)):
\begin{subequations}  \label{scalar_eq_Poisson}
\begin{alignat}{2}
\Psi - \Phi &=&\,\, \Pi\qquad\qquad &= \cA_T\bar{\Pi}, \label{phi} \\
\left({\bf {\cal L}} - {\cal C}_G^2{\bf D}^2\right)\!\Psi
&=&\,\, \sfrac12\Gamma + \left(\sfrac13{\bf D}^2  + {\cal
H}\cL_A \right)\!\Pi  &= \cA_T\!\left(\sfrac12\bar{\Gamma} +
\!\left(\sfrac13{\bf D}^2 +
{\cal H}\partial_\eta +2\cH'\right)\!\bar{\Pi}\right),     \label{bardeen} \\
({\bf D}^2 + 3K)\Psi &=& \,\, \sfrac12 \Delta \qquad\qquad &=
\sfrac12\cA_T\mathbb{D}, \label{Poisson_P} \\
\partial_{\eta}\Psi+ {\cal H}\Phi &=&\,\, - \sfrac12{V} \qquad\qquad &=
-\sfrac12\cA_T\mathbb{V},  \label{V_P}
\end{alignat}
\end{subequations}
where the second order differential operator ${\bf{\cal L}}$ is defined by
\be\label{factorL_s} {\bf {\cal L}}(\bullet) := {\cal
H}\cL_A\cL_B\left(\frac{\bullet}{{\cal H}} \right), \ee
or equivalently
\be \label{L_s}  {\bf {\cal L}} =
\partial_\eta^2 + 3\left(1 + {\cal C}_G^2\right){\cal
H}\partial_\eta +  {\cal H}^2{\cal B} -(1 + 3\cC_G^2)K, \ee
(UW, equation (56)). For a discussion of these two
systems of governing equations, and the ways in which they differ,
we refer to UW, section 3.2, following equation (56).

\subsection{Linearized conservation equations, without Einstein's field equations}
\label{sec:lin_cons}

As shown in Appendix~\ref{app:derconserved}, linearizing the
conservation law $\bna\!_b T^b\!_a = 0$ leads to the following
gauge-invariant equations:
\begin{subequations}\label{cons_X}
\begin{align}
\partial_\eta(\mathbb{D}[X] - 3\Psi[X]) + {\bf D}^2(\mathbb{V}[X] - {\bf B}[X])
&= - 3{\cal H}\bar{\Gamma},\label{cons0} \\
(\partial_\eta + {\cal H})\mathbb{V}[X] + \Phi[X] + \cC_T^2\mathbb{D}
&= -\bar{\Gamma} - {\bar{\Xi}} ,\label{consi}
\end{align}
where
\be \label{Xi}  {\bar{\Xi}} := \sfrac23({\bf D}^2 +
3K)\bar{\Pi},  \ee
\end{subequations}
and $\mathbb{D}[X], \mathbb{V}[X], \bar{\Gamma}$ and
$\bar{\Pi}$ are defined by equations~\eqref{GT}
and~\eqref{T_i}.

These equations are valid for any choice of temporal gauge
field $X^0$. Referring to~\eqref{indep_X} and~\eqref{indep_X2}
we recognize the three groups of terms on the left side as the
$X$-independent expressions $[{\mathbb D}, \Psi]$, $[{\mathbb
V}, {\bf B}]$ and $[\Phi, {\mathbb V}]$ in \eqref{indep_X}. We
choose $X^0=X^0_{\mc}$ and  $X^0=X^0_{\mathrm p}$ in the first
two, and $X^0=X^0_{\mathrm v}$ in the third, and
use~\eqref{gaugechoices}. Equations~\eqref{cons_X} then assume
the concise form
\begin{subequations}
\begin{align}
\partial_\eta\mathbb{D}_\mathrm{c} + {\bf D}^2\mathbb{V} &= - 3{\cal H}\bar{\Gamma} ,\label{Dcevol}\\
\Phi_\mathrm{v} + \cC_T^2\mathbb{D} &= -\bar{\Gamma} - {\bar{\Xi}},
\end{align}
\end{subequations}
where $\mathbb{D}_\mathrm{c}=\mathbb{D}[X_\mathrm{c}]$,
${\mathbb V} = {\mathbb V}[X_{\mathrm p}]$, $\mathbb{D} =
\mathbb{D}[X_\mathrm{v}]$ and $\Phi_\mathrm{v} =
\Phi[X_\mathrm{v}]$, in accordance with our convention for
labeling gauge invariants. In certain circumstances, the first
equation can be interpreted as a conservation law for
$\mathbb{D}_\mathrm{c}$,  as will be discussed in
section~\ref{sec:cons}. The second equation\footnote{This
equation corresponds to equation (5.20) in Bardeen (1980).}
shows that for a barotropic perfect fluid $\Phi_\mathrm{v}$ is
proportional to $\mathbb{D}$, and is in fact zero for dust.

One would like to use equations~\eqref{cons_X} to obtain a
system of evolution equations for the stress-energy gauge
invariants $ \mathbb{D}[X]$ and $ \mathbb{V}[X]$, for some
choice of the gauge field $X$. This is not possible due to the
presence of the metric variables $\partial_\eta \Psi[X]$ and
$\Phi[X]$. However, in the case that the stress-energy tensor
is the total stress-energy tensor one can use the linearized
Einstein equations to eliminate these terms and achieve the
desired goal, as we will show in
section~\ref{sec:cons_plus_einstein}. On the other hand
equations~\eqref{cons_X} are valid for each (non-interacting)
individual stress-energy tensor of a multi-component source,
and as such they form a convenient starting point for the
derivation of a simple system of governing equations for scalar
perturbations of such a source. We will derive these equations
in section \ref{subsec:multi}.

\subsection{Linearized conservation equations in conjunction\\ with Einstein's field equations}
\label{sec:cons_plus_einstein}

We now use the results of sections~\ref{sec:lin_einst}
and~\ref{sec:lin_cons} to derive a system of governing
equations in the form of a first order (in time) system of
partial differential equations with the stress-energy gauge
invariants as basic variables.

Choose $X^0 = X^0_{\mathrm p}$ in~\eqref{cons_X}, and eliminate
${\mathbb D}_{\mathrm p}$ using ${\mathbb D} = {\mathbb D}_{\mathrm p} -
3\cH {\mathbb V}$, which is obtained from the $X$-independent
invariant $[{\mathbb D}, {\mathbb V}]$ in~\eqref{indep_X}. The resulting equations are
\begin{subequations}\label{conservedcompX2}
\begin{align}
\partial_\eta(\mathbb{D} + 3\cH {\mathbb V}) - 3\partial_\eta \Psi + {\bf D}^2\mathbb{V}
&= - 3{\cal H}\bar{\Gamma},\label{cons02} \\
(\partial_\eta + {\cal H})\mathbb{V} + \Phi + \cC_T^2\mathbb{D}
&= -\bar{\Gamma} -  {\bar{\Xi}},  \label{DV2}
\end{align}
\end{subequations}
using the notation~\eqref{gi_notation}. The
combination~\eqref{cons02} $-$ $3\cH$\eqref{DV2} can be
rearranged to read
\begin{equation}
(\partial_\eta - 3\cH\cC_T^2)\mathbb{D} + ({\bf D}^2+3K)\mathbb{V}
- 3(\partial_\eta\Psi + \cH\Phi + \sfrac12\cA_G\mathbb{V}) = 3\cH{\bar{\Xi}},
\end{equation}
where $\cA_G$ is given by~\eqref{A_G}. If the stress-energy
tensor is the total stress-energy tensor, and if we impose
Einstein's field equation~\eqref{V_P} and the background field
equation $\cA_G=\cA_T$, then the above equation simplifies to
\begin{equation}  \label{DV1}
(\partial_\eta - 3\cH\cC_T^2)\mathbb{D} + ({\bf D}^2+3K)\mathbb{V} = 3\cH{\bar{\Xi}} .
\end{equation}
Equations~\eqref{DV1} and~\eqref{DV2} form a coupled
first order system of evolution equations for $\mathbb{D}$ and
$\mathbb{V}$. However due to the appearance of the metric
potential $\Phi$ the system is not closed. We can remedy this
deficiency by applying the operator ${\bf D}^2 + 3K$
to~\eqref{DV2} and using
\be \label{Bbb_Z} {\mathbb Z} :=  ({\bf D}^2 + 3K){\mathbb V}, \ee
as a new variable to replace ${\mathbb V}$ in the system. On using the Einstein
equations~\eqref{phi} and~\eqref{V_P}, which yield
\be ({\bf D}^2 + 3K)\Phi = ({\bf D}^2 + 3K)(\Psi -
\cA_T\bar{\Pi}) = \sfrac12\cA({\mathbb D} - 3{\bar{\Xi}}),  \ee
equations~\eqref{DV1},~\eqref{Bbb_Z} and~$({\bf D}^2 + 3K)(\eqref{DV2})$ result in
\begin{subequations}\label{DZ_evol1}
\begin{align}
(\partial_\eta - 3\cC^2_T\cH){\mathbb D} + {\mathbb Z} &= 3\cH {\bar{\Xi}}, \label{D_evol1}\\
\left(\partial_\eta  + {\cal H}\right) {\mathbb Z}  +
\left(\sfrac12\cA + {\cal C}_T^2({\bf D}^2 + 3K)\right)\!{\mathbb D}  &=
- ({\bf D}^2 + 3K)(\bar{\Gamma} + {\bar{\Xi}}) + \sfrac32\cA{\bar{\Xi}},  \label{Z_evol1}
\end{align}
\end{subequations}
where $\cA\equiv\cA_G=\cA_T$. For the reader's convenience we note
that the variables in these equations are defined by
equations~\eqref{bf_T_i},~\eqref{bf_T},~\eqref{Xi}
and~\eqref{Bbb_Z}.

Equations~\eqref{DZ_evol1} constitute one of the main results
of this paper. They form a coupled system of first order (in
time) partial differential equations for $({\mathbb D}, {\mathbb Z})$, assuming that the
stress-energy terms $\bar{\Gamma}$ and ${\bar{\Xi}}$ are given.
They determine the behaviour of the scalar mode of linear
perturbations of an FL cosmology with arbitrary stress-energy
content. The structure of this system is similar to the
structure of the system of evolution equations~\eqref{bfB_evol}
and~\eqref{Phicurv_evol} for the uniform curvature  metric
gauge invariants ${\bf A}$ and ${\bf B}$, and can be derived
from them as follows. First use~\eqref{V_F} to express ${\bf
A}$ in terms of ${\mathbb V}$. Then apply the operator ${\bf D}^2
+ 3K$ to both equations and use~\eqref{Poisson_C} to express
$({\bf D}^2 + 3K){\bf B}$ in terms of $\cA_T {\mathbb D}$, after which
some obvious manipulations lead to equations~\eqref{DZ_evol1}.

\subsubsection*{The evolution equation for ${\mathbb D}$}

By eliminating ${\mathbb Z}$ from equations~\eqref{DZ_evol1} one
can obtain a second order evolution equation for the
gauge-invariant density perturbation ${\mathbb D}$. We apply the
operator $\partial_\eta + \cH$ to the first of
equations~\eqref{DZ_evol1} and use the second equation to
eliminate ${\mathbb Z}$. The resulting equation can be written in the form
\begin{subequations}
\be  \label{evol_D}  \left({\bf {\cal L}}_{\cal D} - \cC^2 {\bf
D}^2\right)\!{\mathbb D} = 2({\bf D}^2 + 3K)\left(\sfrac12\bar{\Gamma} + (\sfrac13
{\bf D}^2 + \cH \partial_\eta +2\cH' ){\bar{\Pi}}\right) , \ee
where
\be\label{L_D}  {\bf {\cal L}}_{\cal D} :=
\partial_\eta^2 + \left(1 - 3{\cC}^2\right)\!{\cH}\partial_\eta +
(1 - 3{\cC}^2 )\cH' - (1 + 3\cC^2)(\cH^2 +K) - 3(\cC^2)' \cH .
\ee
\end{subequations}
Equation~\eqref{evol_D} can also be derived from the governing
equations~\eqref{scalar_eq_Poisson} in Poisson form. We apply
${\bf D}^2 + 3K$ to~\eqref{bardeen} and use~\eqref{Poisson_P}
to relate $({\bf D}^2 + 3K)\Psi$ to ${\mathbb D}$. By comparing
the resulting evolution equation with~\eqref{evol_D}, we can
conclude that the operator ${\bf {\cal L}}_{\cal D}$ is related to
the operator ${\bf {\cal L}}$ according to
\be\label{Lbullet_A} {\bf {\cal L}}(\cA \,\bullet) = \cA {\bf
{\cal L}}_{\cal D}(\bullet).  \ee
This result can also be verified by direct calculation.

Equation~\eqref{evol_D} is a second order linear partial
differential equation for ${\mathbb D}$, assuming that the
stress-energy terms ${\bar \Gamma}$ and ${\bar \Pi}$ are given.
It differs from other related equations in the literature, for
example, Ellis {\it et al} (1990)\,(see their equation (48),
with the coefficients given by equations (19) and (20)), and
Hwang and Noh (1999) (see their equation (45)), since we have
defined  ${\mathbb D}$ by normalizing with ${\cal M}^2 =
{}^{(0)}\!\rho + {}^{(0)}\!p$, while the usual practice is to
use ${}^{(0)}\!\rho$. If the source is a perfect fluid with a
linear equation of state and a cosmological constant, {\it
i.e.}
\be \rho = \rho_m + \Lambda, \quad  p = p_m - \Lambda, \quad p_m=w\rho_m,  \ee
then the right side of~\eqref{evol_D} is zero, and one can use
Einstein's equations in the background model (UW equations
(41)) to write the expression~\eqref{L_D} in the
form\footnote{Since  $w=constant$ we have $\cC^2 = w$.}
\be\label{L_A2}  {\bf {\cal L}}_{\cal D} =
\partial_{\eta}\!^2 +(1 - 3w){\cH}\partial_\eta -
\sfrac12 [ (1+3w)(1-w)\rho_m + 4w\Lambda]a^2.   \ee
In this case~\eqref{evol_D} is compatible with those equations
cited above.

\subsection{Governing equations for a multi-component source}
\label{subsec:multi}

The governing equations for  a perturbed FL cosmology with a
multi-component source were first derived by Kodama and Sasaki
(1984), and subsequently considered by various authors
including Hwang (1991), Dunsby {\it et al} (1992) and Durrer
(2008). These authors, as is customary, use normalization with
${}^{(0)}\!\rho$ when defining the density perturbation. We
have found that the derivation and the form of the governing
equations is significantly simpler if one uses ${\cal
M}$-normalization, as introduced in
section~\ref{sec:stress-energy}. In this section we thus give a
brief derivation of the relevant equations.

We consider a multi-component source with $n$ separate
stress-energy tensors denoted by $_{A}T^a\!_b,$ with $A =
1,\dots,n$,  which sum to form the total stress-energy tensor:
\be T^a\!_b =  \nsum _{A}T^a\!_b.     \label{T_sum} \ee
For simplicity we assume that the individual components are
\emph{non-interacting}. As shown in
Appendix~\ref{app:derconserved}, linearizing the conservation
equation $\bna\!_b\, {}_{A}T^b\!_a = 0$ for an arbitrary
component labeled $A$ leads to the following
equations\footnote{In these equations $\cC^2_A$ denotes the
value of $\cC^2_T$ for the component labeled $A$. We have
dropped the subscript $T$ for convenience.}:
\begin{subequations} \label{conserv_A}
\begin{align}
\partial_\eta(\mathbb{D}_A[X] - 3\Psi[X]) + {\bf D}^2(\mathbb{V}_A[X] - {\bf B}[X])
&= - 3{\cal H}\bar{\Gamma}_A, \label{conserv1_A}\\
(\partial_\eta + {\cal H})\mathbb{V}_A[X] + \Phi[X] + \cC_A^2\mathbb{D}_A
&= -\bar{\Gamma}_A - {\bar{\Xi}}_A, \label{conserv2_A}
\end{align}
\end{subequations}
where
\be   {\bar{\Xi}}_A := \sfrac23({\bf D}^2 + 3K)\bar{\Pi}_A.
\ee
We assume that the gauge field $X$ does not depend on the
labeling index. A quantity ${\mathbb F}_A$ associated with
$_{A}T^a\!_b$ that satisfies
\be \nsum{{\mathcal B}_A {\mathbb F}_A} = {\mathbb F},   \ee
where $ {\mathbb F}$ is the corresponding quantity associated
with $T^a\!_b$, will be called \emph{additive}. Here the
coefficients ${\mathcal B}_A$ are defined by
\be {\mathcal B}_A:=\frac{\cA_A}{\cA}, \quad \text{and satisfy}
\quad \nsum{ {\mathcal B}_A} = 1. \ee
We note that ${\mathbb D}_A[X]$,  ${\mathbb V}_A[X]$, ${\mathbb
D}_A, \Xi_A$, ${\bar \Gamma}_A + \cC^2 {\mathbb D}_A[X]$ and
$\cC_A^2$ are additive\footnote{One can check the consistency
of~\eqref{conserv_A} by showing that $\nsum {\mathcal
B}_A\eqref{conserv_A} =\eqref{cons_X}$. Note that ${\mathbb
D}_A = {\mathbb D}_A[X] - 3\cH {\mathbb V}_A[X].$}.

In order to obtain a closed system of evolution equations we
introduce the "difference variables"
\be  {\mathbb D}_{AB} := {\mathbb D}_A[X] - {\mathbb D}_B[X],
\qquad {\mathbb V}_{AB} := {\mathbb V}_A[X] - {\mathbb V}_B[X].
\ee
It follows from~\eqref{DX},~\eqref{TiX},~\eqref{hybrid_T}
and~\eqref{T_00} that ${\mathbb D}_{AB}$ and ${\mathbb V}_{AB}$ are
$X$-independent. In order to obtain evolution equations for
${\mathbb D}_{AB}$ and ${\mathbb V}_{AB}$ we form the difference of
two copies of equations~\eqref{conserv_A}, labeled $A$ and $B$.
%
In this way we obtain the following equations\footnote{The only
calculation involves showing that $\cC_A^2\mathbb{D}_A -
\cC_B^2\mathbb{D}_B = {\mathbb K}_{AB} +  (\cC^2_A -
\cC^2_B){\mathbb D}$, which follows by writing ${\mathbb D}_A =
{\mathbb D} + {\sum_{C=1}^n { {\mathcal B}_C({\mathbb D}_A -
{\mathbb D}_C) }}$, and a similar expression for ${\mathbb D}_B
$.}:
\begin{subequations} \label{conserv_AB}
\begin{align}
\partial_\eta {\mathbb D}_{AB} + {\bf D}^2 {\mathbb V}_{AB} &=
-3\cH {\bar \Gamma}_{AB} ,\\
(\partial_\eta + \cH) {\mathbb V}_{AB} \, + {\mathbb K}_{AB} + (\cC_A^2 - \cC_B^2) {\mathbb D}
&= - {\bar \Gamma}_{AB} - {\bar \Xi}_{AB},
\end{align}
where
\begin{align}
{\mathbb K}_{AB} &:= {\sum_{C=1}^n { {\mathcal B}_C
\left(\cC_A^2({\mathbb D}_{AC} -3\cH {\mathbb V}_{AC})  -
\cC_B^2 ({\mathbb D}_{BC} -3\cH {\mathbb V}_{BC})\right)}}, \\
{\bar \Gamma}_{AB} &:=
{\bar \Gamma}_A - {\bar \Gamma}_B, \qquad  {\bar \Xi}_{AB} :=
{\bar \Xi}_A - {\bar \Xi}_B,
\end{align}
\end{subequations}
Equations~\eqref{conserv_AB} do not form a closed system for
${\mathbb D}_{AB}$ and ${\mathbb V}_{AB}$, however, due to the
appearance of the total intrinsic matter gauge invariant
${\mathbb D}$. The evolution equation~\eqref{evol_D} for
${\mathbb D}$ contains ${\bar \Gamma}$ and ${\bar \Xi}$ as
source terms\footnote{Note that the term $({\bf D}^2 + 3K){\bar
\Pi}$  on the right side of~\eqref{evol_D} can be replaced by
${\bar \Xi}$ on account of~\eqref{Xi}.}, which have to be
expressed in terms of  ${\bar \Gamma}_{A}$, ${\bar \Xi}_{A}$
and ${\mathbb D}_{AB}$. The term  ${\bar \Gamma}$ is not
additive, whereas ${\bar \Gamma} + \cC^2 {\mathbb D}[X]$ is,
{\it i.e.}
\be  \label{gamma_sum}  {\bar \Gamma} + \cC^2 {\mathbb D}[X] =
\nsum {\mathcal B}_A({\bar \Gamma}_A  + \cC^2_A {\mathbb D}_A[X]). \ee
It follows that ${\bar \Gamma}$ can be
expressed\footnote{Substitute ${\mathbb D}_A[X] = {\mathbb
D}[X] + {\sum_{C=1}^n { {\mathcal B}_C{\mathbb D}_{AC} }} $
in~\eqref{gamma_sum} and use $\nsum {\cal B}_A {\cC}^2_A =
\cC^2$.} as a sum involving  ${\bar \Gamma}_A$ and ${\mathbb
D}_{AB}$:
\begin{subequations}  \label{sum}
\be \label{sum_gamma} {\bar \Gamma} =
\nsum {\mathcal B}_A {\bar \Gamma}_A +
\sfrac12 {\sum_{A,B=1}^n {{\mathcal B}_A{\mathcal B}_B
(\cC_A^2 - \cC_B^2) {\mathbb D}_{AB}}}. \ee
On the other hand, the term ${\bar \Xi}$ is additive:
\be \label{sum_xi} {\bar \Xi} = \nsum {\mathcal B}_A{\bar \Xi}_A. \ee
\end{subequations}

In conclusion, equations~\eqref{conserv_AB} and the evolution
equation~\eqref{evol_D} for ${\mathbb D}$, in conjunction with
equations~\eqref{sum}, form a closed system for ${\mathbb
D}_{AB}, {\mathbb V}_{AB}$ and ${\mathbb D}$ that describes the
scalar mode of a perturbed FL cosmology with a multi-component
source\footnote {For the reader's convenience we give the
equation numbers in the previously mentioned references that
correspond to our equations~\eqref{conserv_AB}: Kodama and
Sasaki (1984), (5.59)-(5.60), Hwang (1991), (37)-(38), Bruni
{\it et al} (1992b), (91)-(92), Dunsby {\it et al} (1992),
(86)-(87), and Durrer (2008), (2.136)-(2.137).}. We note that
all gauge invariants in these equations are $X$-independent.

\subsection{Gauge-fixing versus gauge-invariance}

To conclude this section we comment briefly on the two points
of view as regards formulating the governing equations for
scalar perturbations. In the gauge-fixing approach one chooses
a gauge {\it ab initio}, in which case all the gauge invariants
that appear in the governing equations are associated with this
particular gauge. Recent references that use this traditional
approach are Mukhanov (2005) and Weinberg (2008). In contrast,
one can work with a variety of gauge invariants in the spirit
of Bardeen (1980), in which case one has the flexibility to use
gauge invariants associated with different gauges in
formulating the governing equations. A notable example is the
generalized Poisson equation~\eqref{Poisson_P}, which relates
the Bardeen potential  $\Psi\equiv\Psi[X_{\mathrm p}]$ to the
matter density gauge invariant ${\mathbb D}\equiv{\mathbb
D}[X_{\mathrm v}]$. A variety of other examples occur in this
paper, including the system of equations \eqref{DZ_evol1} for
${\mathbb D}_{\mathrm v}$ and ${\mathbb Z}_{\mathrm p}$,
equations~\eqref{Poisson_C} and~\eqref{V_F} in the uniform
curvature formulation, and the conservation
law~\eqref{zeta_evol1}.

\section{Conserved quantities}
\label{sec:cons}

Two "conserved quantities" that are associated with scalar
perturbations of FL  have been defined in the literature. These
quantities, often denoted by $\zeta$, satisfy an evolution
equation of the form
\be \label{cons} \partial_\eta \zeta_{ \,\bullet} = {\bf D}^2
{\bf C}_{\,\bullet} + \cH \bar{\Gamma},  \ee
where ${\bf C}_{\,\bullet}$ is an expression involving the
primary gauge invariants such as $\Psi$ or ${\mathbb V}$ and the
background variables. This equation is referred to as a
"conservation law", since if spatial derivatives are negligible
("perturbations outside the horizon") and if $\bar{\Gamma}$ is
zero or negligible in some epoch, then~\eqref{cons} is
approximated by $\partial_{\eta}\zeta_{ \,\bullet} = 0$, {\it
i.e.} $\zeta_{ \,\bullet}$ is approximately constant in time
during that epoch.

Two of the evolution equations that we have presented in
section~\ref{sec:equations}, namely equations~\eqref{Dcevol}
and~\eqref{Phicurv_evol}, can be written in the
form~\eqref{cons} simply by multiplying by a suitable factor.
The conserved quantity can then be identified by inspection. We
consider each in turn.

\subsubsection*{The conserved quantity $\zeta_\rho$}

The evolution equation~\eqref{Dcevol} for ${\mathbb D}_\mathrm{c}$
has the form of a conservation law~\eqref{cons}. We multiply by
the numerical factor $-\frac13$ to agree with current convention,
and comparison with~\eqref{cons} leads to the following
definition of the first conserved quantity:
\be \zeta_{\rho} := -\sfrac13 {\mathbb D}_{\mc}.    \label{zeta_rho}
\ee
Equation~\eqref{Dcevol}  assumes the form
\be  \label{zeta_evol1}
\partial_\eta \zeta_{\rho} = \sfrac13 {\bf D}^2 {\mathbb V} + \cH\bar{\Gamma}.
\ee
Our motivation for using the notation $\zeta_\rho$ is that
on account of  equation~\eqref{DcPsirho} we have
\be \label{zeta_rho2} \zeta_\rho = \Psi\!_\rho.  \ee
This conservation law was apparently first given in a form
closely related to the above by Wands \emph{et al}
(2000),\footnote{See their equations (8) and (9). Their
evolution equation (18) corresponds to our
equation~\eqref{zeta_evol1}, although their equation is not in
a manifestly gauge-invariant form, and uses clock time rather
than conformal time. See also Malik and Wands (2009), equations
(7.61), (7.62) and (8.35), which are somewhat closer in form to
our equations.} who emphasized that it depends only on the
conservation equation for the stress-energy tensor, \emph{i.e.}
it is independent of Einstein's equations. They denoted
our ${\zeta_\rho}$ by ${\zeta}$ and because
of~\eqref{zeta_rho2} they referred to it as "the curvature
perturbation on uniform density surfaces." The quantity
$\zeta_\rho$ has its origins in the paper Bardeen, Steinhardt
and Turner (1983), and was further studied from a different
point of view by Brandenberger and Khan (1984)\footnote {It is
not immediately obvious that the expressions given in these
papers (equations (2.43) and (2.45) in Bardeen {\it et al}
(1983) and equations (2.11) and (2.12) in Brandenberger and
Khan (1984)) agree with our expressions.}.

A major step in understanding the significance of
$\zeta_{\rho}$ was taken by Langlois and Vernizzi (2005).
Motivated by the $1 + 3$ covariant approach,  they showed that
this quantity could be obtained as the linearization of an
exact nonlinear evolution equation for a certain covector. This
approach enabled them to extend the definition of
$\zeta_{\rho}$ to second-order (and higher order)
perturbations. We refer to their equations (20) and (25) for
the general situation and note that their equations (41) and
(42) correspond to our equations~\eqref{zeta_rho2}
and~\eqref{zeta_evol1}.

\subsubsection*{The conserved quantity $\zeta_{\mathrm v}$}

The second conservation equation arises from the evolution
equation~\eqref{Phicurv_evol} for ${\bf A}\equiv\Phi_{\mc}$.
Using~\eqref{cB} we can write the differential operator ${\bf{\cal L}}_A$
in the form:
\be \cH{\bf {\cal L}}_A(\bullet) = \frac{\cA_G}{\cH}
\partial_\eta\left(\frac{\cH^2}{\cA_G}\bullet\right).  \ee
Thus on multiplying~\eqref{Phicurv_evol} by $2\cH/\cA_G$ we can
write it in the form of a conservation law\footnote{Throughout
the remaining discussion about conserved quantities we use the
background Einstein equations and hence $\cA_G=\cA_T=\cA$
and $\cC_G^2=\cC_T^2=\cC^2$.}
\be \label{Phi_c_evol}  \partial_\eta \left(\frac{2{\cal
H}^2}{{\cal A}} \Phi_{\mathrm c} \right) =  2{\cal H}{\bf D}^2
\left( {{\cal C}^2}\frac{\Psi} {\cal A}  + \sfrac13\bar{\Pi}
\right) + {\cal H}\bar{\Gamma}, \ee
where we have chosen to replace ${\bf B}$ by $\Psi$, using the
relation $\Psi = - \cH {\bf B}$. Comparing~\eqref{Phi_c_evol}
with~\eqref{cons} leads to the following definition of the
second conserved quantity:
\be \zeta_{\mathrm v} :=  \frac{{2\cal H}^2}{ {\cal A}}
\Phi_{\mathrm c}.   \label{zeta_v} \ee
Equation~\eqref{Phi_c_evol}, which is equivalent
to~\eqref{Phicurv_evol}, is the conservation equation for
$\zeta_{\mathrm v}$. An immediate consequence of this equation
is that \emph{if the source is pressure-free matter plus
possibly a cosmological constant, then $\zeta_{\mathrm v}$ is
constant in time.}

We can derive an alternate expression for the  conserved
quantity $\zeta_{\mathrm v}$ as follows. Using the relation
$\Psi = -\cH {\bf B}$ and the definition~\eqref{zeta_v}, the
velocity equation~\eqref{V_F} can be written in the form:
\be \label{zeta_v1}  \zeta_{\mathrm v} = \left( 1 -\frac{2K}{\cA}
\right)\Psi - \cH {\mathbb V} . \ee
We use the $X$-independent gauge invariant $[\Psi,{\mathbb V}]$
in~\eqref{indep_X} to obtain
\be \Psi_\mathrm{v} = \Psi - \cH\mathbb{V} , \ee
which when inserted in~\eqref{zeta_v1} leads to
\be  \label{zeta_v2}  \zeta_{\mathrm v}  = \Psi_{\mv}  -
\frac{2K}{\cA}\Psi . \ee
The quantity $\zeta_{\mathrm v}$ is most commonly used when the
background curvature is zero ($K = 0$) in which case
\be \label{zeta_v3}  \zeta_{\mathrm v} |_{K=0} = \Psi_{\mv}. \ee
This expression motivates our use of the notation
$\zeta_{\mathrm v}$. Malik and Wands (2009) use the notation
$\cal R$ for $\zeta_{\mathrm v}$ in the case $K=0$ (see
equation (7.46)), and refer to it as "the curvature
perturbation in the comoving gauge." This quantity has its
origin in the paper Bardeen (1980) (see equations (5.19) and
(5.21)).

There is another commonly used expression for $\zeta_{\mathrm
v}$, in terms of the Bardeen potential $\Psi$ and its time
derivative, which we can quickly derive. On
solving~\eqref{bfB_evol} for ${\bf A}\equiv \Phi_{\mc}$, the
definition~\eqref{zeta_v} yields
\be \zeta_{\mathrm v} = \frac{2\cH^2}{\cA_G} \left({\bf{\cal
L}}_B\left(\frac{\Psi}{\cH}\right) - \Pi \right), \ee
where we used $\cH {\bf B} = - \Psi$ and where ${\bf{\cal
L}}_B$ is defined by~\eqref{L_B}. We expand the operator and
use~\eqref{A_G} to express $\cH'$ in terms of $\cA_G$. On
specializing to flat FL ({\it i.e.} $K=0$) and setting $\Pi=0$
we obtain the expression
\be  \label{zeta_familiar} \zeta_{\mathrm v}|_{K=0} = \Psi +
\frac{2}{3(1+w)}\left(\frac{1}{\cH} \Psi' + \Psi\right), \ee
where $w:= {}^{(0)}\!p/^{(0)}\!\rho$. Here we have used the
background field equation $\cA_G = \cA_T$, and the fact that
$\cA_T = 3(1 + w)\cH^2$ when $K=0$, as follows from UW (see
equations (41a) and (42a)). The familiar
expression~\eqref{zeta_familiar} can be found, for example, in
Mukhanov {\it et al} (1992), equation (5.23), and Mukhanov
(2005), equation (7.73).

We conclude this section by showing that $\zeta_\rho$ and
$\zeta_{\mathrm v}$, despite their different origins, are
closely related. On account of~\eqref{zeta_rho2}
and~\eqref{zeta_v2}
\be \zeta_{\rho} - \zeta_{\mathrm v} = \Psi_\rho -
\Psi_{\mathrm v} + \frac{2K}{\cA}\Psi =  -\sfrac13\mathbb{D} +
\frac{2K}{\cA}\Psi, \ee
where we have used the $X$-independent invariant $[{\mathbb
D},\Psi]$ in~\eqref{indep_X} to obtain the second equality. The
generalized Poisson equation~\eqref{Poisson_P} can be used to
eliminate $\mathbb{D}$ yielding
\be \zeta_{\rho} - \zeta_{\mathrm v} = - \frac{2} {3\cA} {\bf
D}^2 \Psi.      \label{3.3} \ee
This equation suggests that \emph{if spatial derivatives are
negligible in some epoch, then} $\zeta_{\rho} \thickapprox
\zeta_{\mathrm v}$ in that epoch.

\subsubsection*{A coupled system for $(\zeta_{\mathrm v}, \Psi)$}

The conservation equation~\eqref{Phi_c_evol} for
$\zeta_{\mathrm v}$ is one of the linearized Einstein equations
in uniform curvature form, namely, the evolution equation for
${\bf A} = \Phi_{\mc}$. It is helpful to also write the
evolution equation~\eqref{bfB_evol} for ${\bf B}$ in terms of
$\zeta_{\mathrm v}$, replacing ${\bf B}$ by $\Psi$ using the
relation $\Psi = - \cH {\bf B}$. The resulting pair of
equations has the following form:
\begin{subequations} \label{zeta_psi}
\begin{align} \partial_\eta\left(\frac{x^2}{\cH}\Psi \right) -
{\mathcal F}\zeta_{\mathrm v} &= x^2 \cA \bar{\Pi}, \\
\partial_\eta \zeta_{\mathrm v} -  \frac{\cC^2}{\mathcal F}
{\bf D}^2\left(\frac{x^2}{\cH} \Psi\right) &=
\cH\!\left(\bar{\Gamma} + \sfrac23{\bf D}^2 \bar{\Pi}
\right),
\end{align}
\end{subequations}
where $x := a/a_*$ is the dimensionless scale factor, with
$a_*$ being the scale factor at some reference time, and
\be {\mathcal F}:= \frac{x^2 \cA}{2\cH^2}. \ee
The system of equations~\eqref{zeta_psi} is a particularly
useful form of the governing equations for scalar perturbations
of FL. The second equation is the "conservation equation" for
$\zeta_{\mathrm v}$, while the first equation enables one to
express the Bardeen potential $\Psi$ as a quadrature, in
situations in which $\zeta_{\mathrm v}$ is a temporal constant
(or can be treated as such) and $\bar{\Pi}$ is negligible:
\be \Psi(\eta, x^i) =  \frac{\mathcal{H} }{x^2} \left(
\zeta_{\mathrm v}(x^i) \int_0 ^{\eta} {\mathcal
F}d\overline{\eta} + C_{-}(x^i)\right)  .  \label{psi} \ee
In particular this equation gives the {\it exact} solution
where the source is pressure-free matter and, possibly, a
cosmological constant (since then $\zeta_{\mathrm v}$ is a
temporal constant), and the {\it approximate} solution in the
long wavelength limit, in both cases without restriction on the
background spatial curvature $K$.

\section{The $1+3$ gauge-invariant approach}
\label{sec:1+3}

In this section we first give a concise derivation of the
governing equations for linear scalar perturbations in the
$1+3$ approach, combining the formulation of Bruni, Dunsby and
Ellis (1992a) (hereafter referred to as BDE)\footnote{BDE give a
comprehensive account of the linearization of the full system
of equations in the $1+3$ formalism (Ricci and Bianchi identities,
and stress-energy conservation equations ).
We are concerned only with a limited subset of these
equations.} with our overall strategy of creating dimensionless
quantities by normalizing with $\cal M$ and $a$. We then
introduce the dependence of the $1+3$ variables and
differential operators on a perturbation parameter $\epsilon$,
which enables us to relate the variables and governing
equations of the two approaches in a precise manner. In this
way we set the stage for extending the $1+3$ approach to second
order.

The $1+3$ gauge-invariant approach to cosmological
perturbations is based on choosing a preferred unit timelike
vector field $u^a$ and decomposing the stress-energy tensor
relative to this vector field:
\be \label{T_ab}  T^a\!_b = (\rho + p)u^a u_b + p\delta ^a\!_b
+ (q^au_b + u^aq_b) + \pi^a\!_b,   \ee
where
\be u^a q_b = 0,  \quad  \pi^a\!_a = 0, \quad u_a \pi^a\!_b =
0. \ee
One distinguishes between physical and geometrical quantities
which are non-zero in the background spacetime, namely $\rho,
p$ and the Hubble scalar $H$ associated with $u^a$, and
quantities which are zero in the background, such as the
stress-energy quantities $q_a, \pi^a\!_b$, the shear of the
preferred congruence, the Weyl curvature tensor and the spatial
gradients of $\rho, p$ and $H$ orthogonal to $u^a$. The $1+3$
approach focusses on the latter quantities, which are
gauge-invariant on account of the Stewart-Walker Lemma.

\subsection{Evolution equations}

The spatial gradients of $\rho$ and $H$ describe the
perturbation in a gauge-invariant way at the linear level.
However, in order to extract the scalar mode of the
perturbation it is necessary to form scalar quantities. We thus
take the spatial divergence of these spatial gradients and form
the {\it dimensionless spatial Laplacian}:
\be \label{DZ_defn} D:= (a^2\,{^{(3)}{\bna}^2}){\cal M}^2\rho,
\qquad   Z:= 3(a^2\,{^{(3)}{\bna}^2})aH,   \ee
where
\begin{subequations} \label{laplacian}
\be   {}^{(3)}{\bna}\!_a := h_a\!^b {\bna}\!_b, \qquad
^{(3)}{\bna}^2:= g^{ab} \,{^{(3)}{\bna}\!_a}
{^{(3)}{\bna}\!_b}, \ee
and
\be h_a\!^b := \delta_a\!^b  + u_a u^b. \ee
\end{subequations}
Note that in introducing dimensionless variables we are
normalizing the energy density $\rho$ with ${\cal M}^2$ and the
geometric quantity $H$ with the background scale factor $a$.
Likewise we normalize the geometric operator $^{(3)}{\bna}^2$
with $a^2$.

The governing equations for the scalar mode  take the form of a
coupled system of first order (in time) partial differential
equations for $D$ and $Z$. These equations arise from the
energy conservation equation (the exact evolution equation for
$\rho$) and the Raychaudhuri equation (the exact evolution
equation for $H$). To derive the governing equations one simply
applies the differential operator $a^2\,{^{(3)}{\bna}^2}$ to
the linearized versions of the evolution equations for $\rho$
and $H$, which are obtained by dropping products of first order
quantites.

The linearized evolution equations for $D$ and $Z$, derived in
Appendix \ref{app:1+3}, are as follows\footnote{Equations
equivalent to~\eqref{1+3_evol} have been derived by Woszczyna
and Kulak (1989) in the case of a barotropic perfect fluid,
using the method of Appendix \ref{app:1+3} but with different
normalization factors (see their equations (11) and (18)).
Their variables $\Delta\epsilon$ and $\Delta\theta$ are related
to $D$ and $Z$ according to $D=a^2{\cal M}^2 \Delta\epsilon,
\quad Z=a^3 \Delta\theta$.}:
\begin{subequations}  \label{1+3_evol}
\begin{align}
D' - 3\cH {\cal C}_T^2 D + Z
&= 3\cH(\tilde{\Pi} + \tilde{ \Upsilon}) -  {\tilde {\bf D}}^2 \tilde {Q}, \label{D_evol}\\
\!\!\!\!\!\! Z' + {\cal H}Z  + \left[\sfrac12\cA + {\cal C}_T^2({\tilde {\bf D}}^2 + 3K)\right]\!\!D  &=
- ( {\tilde {\bf D}}^2 + 3K)(\tilde{\Gamma} + \tilde{\Pi} + \tilde{\Upsilon})
+ \sfrac32\cA(\tilde{\Pi} + \tilde{\Upsilon}),\label{Z_evol}
\end{align}
\end{subequations}
In these equations the dimensionless operators $'$  and
${\tilde {\bf D}}^2$ are defined by
\be \label{operators1}  A': = au^a {\bna}\!_a A, \quad {\tilde
{\bf D}}^2 A :=  a^2\,{}^{(3)}{\bna}^2 A, \ee
where $A$ is a scalar. The source terms $\tilde {\Pi}, \tilde
{Q}$ and $\tilde{\Upsilon}$ are first order dimensionless scalars formed by
taking the spatial divergence of $q^a$ and $\pi^a\!_b$ after
normalizing with ${\cal M}^2$:
\be \label{tildeQ} \tilde{Q}:= \tilde{\bf D}^a({\cal M}^2 q_a),
\qquad \tilde{\Pi}:= {\tilde {\bf D}}_a {\tilde {\bf D}}^b
({\cal M}^2 \pi ^a\!_b),  \qquad \tilde{\Upsilon}: = {\tilde
Q}' - (3{\cal C}_T^2 - 1) \cH\tilde {Q}, \ee
where
\be \label{operators4} {\tilde {\bf D}}_a:= a{}^{(3)}{\bna}\!_a.     \ee
The entropy perturbation ${\tilde \Gamma}$ is given
by\footnote{See BDE equations (27), (28) and (40). Note that
${\tilde \Gamma} = \frac{w}{1+w}{\cal E}$.}
\be \label{tildeGamma} {\tilde \Gamma} = P - {\cal C}_T^2 D,  \ee
where
\be P := {\tilde {\bf D}}^2 ({\cal M}^2 p).  \ee

We conclude this section by relating our approach to that of
BDE\footnote{BDE do not incorporate the cosmological constant
into the stress-energy tensor as we do. In making a comparison
we have to write $\rho=\rho_m +\Lambda$, $p=p_m-\Lambda$, with
$w= p_m/\rho_m$ not necessarily constant.}. The variables $D$
and $Z$ differ from those introduced by BDE as regards the
normalization of $\rho_m$ and $H$. Specifically, BDE define
\be \label{BDE1}  \Delta:=  a^{(3)}{\bna}^a
\left(\frac{a^{(3)}{\bna}\!_a\,\rho_m}{\rho_m}\right) , \quad {\cal
Z}: = 3a^{(3)}{\bna}^a\left(a^{(3)}{\bna}\!_a\,H\right). \ee
At the linear level, the factor $\rho_m$ in the denominator can
be replaced by ${}^{(0)}\!\rho_m$. In the $1+3$ approach the
scale factor is usually defined using the Hubble scalar $H$ of
the preferred congruence $u^a$, according to $(u^a{\bna}\!_a\,
a)/a= H$, in which case $^{(3)}{\bna}^a a\neq 0$. At the linear
level, however, the factor of $a$ can be taken outside
${}^{(3)}{\bna}^a$, since the term $^{(3)}{\bna}^a a$ appears
as a product with ${}^{(3)}{\bna}^a \rho_m$ or $^{(3)}{\bna}\!_a
H$, and hence can be dropped. To avoid this complication we
have chosen the scale factor $a$ to be the scale factor in the
background model so that $^{(3)}{\bna}^a a= 0$. This choice
also facilitates the link with the metric-based approach. In view of
these remarks, the BDE variables~\eqref{BDE1} can be written in
the form:
\be \Delta = \frac{1}{{}^{(0)}\!\rho_m} {\tilde {\bf D}}^2
\rho_m, \quad {\mathcal Z} = 3{\tilde {\bf D}}^2 H, \ee
which implies that the BDE variables are related
to ours according to\footnote
{Note that $\rho+p=\rho_m+p_m=(1+w)\rho_m$, and that
$ {\tilde {\bf D}}^2 \rho_m= {\tilde {\bf D}}^2 \rho$.}
\be  \Delta=(1+w)D, \quad  a{\mathcal Z}= Z,   \ee
since $1/{\cal M}^2 = (1+w){}^{(0)}\!\rho_m.$ Our evolution
equations~\eqref{1+3_evol} for $D$ and $Z$ are equivalent to
equations (68) and (69) for $\Delta$ and ${\cal Z}$ in BDE, but
are simpler in form due to our use of dimensionless variables,
in particular our use of ${\cal M}$-normalization\footnote {Our
variables $(\tilde{Q}, \tilde{\Pi}, \tilde{\Upsilon})$ are
related to those used in the above reference according to
\newline $\Psi_{BDE} = \tilde {Q}, \quad a\Pi_{BDE} = {\tilde
\Pi},  \quad aF_{BDE} = {\tilde \Upsilon}$.}.

\subsection{Relation with the metric-based approach}

Equations~\eqref{1+3_evol} are closely related to the governing
equations in the form~\eqref{DZ_evol1} for the variables ${\mathbb
D}$ and ${\mathbb Z},$ that arise in the metric-based approach. Indeed a
formal similarity is obvious on inspection. However, in order
to relate the two sets of equations we have to regard each of
the variables $D, Z, {\tilde Q}, {\tilde \Pi}$ and ${\tilde
\Gamma}$ in~\eqref{1+3_evol} as being a function of the
perturbation parameter $\epsilon$, which can be expanded in a
power series of the form:
\be  \label{D_eps}  F(\epsilon) = {}^{(0)}\!F + \epsilon\,
{}^{(1)}\!F +\dots\,\,.  \ee
Since each variable is zero in the background
 we have $^{(0)}\!F = 0$,
while $^{(1)}\!F$ is the linear perturbation of
$F$. We also need to consider the dependence of the
differential operators on $\epsilon$, which is as follows:
\be \label{operators2} (A')(\epsilon) =
au^a(\epsilon)^{\epsilon}{\bna}\!_a A(\epsilon), \qquad
({\tilde {\bf D}}^2 A)(\epsilon) =
a^2h^{ab}(\epsilon)^{\epsilon}{\bna}\!_a
\,^{\epsilon}{\bna}\!_bA(\epsilon)  \ee
Assuming that $A$ is a scalar such that $ {}^{(0)}\!A = 0,$ it
follows that
\be \label{operators3}  {}^{(1)}\!(A') = \partial_\eta
{}^{(1)}\!A, \qquad {}^{(1)}\!({\tilde {\bf D}}^2 A) = {\bf
D}^2\,{}^{(1)}\!A,  \ee
as is shown in Appendix~\ref{app:1+3}.

If we now differentiate equations~\eqref{1+3_evol} with respect
to $\epsilon$ and set $\epsilon=0$ the resulting equations have
precisely the same form but with each variable  replaced by its
linear perturbation and each differential operator replaced by
the corresponding zeroth order operator. We finally have to do
a calculation using~\eqref{D_eps} to specifically relate the
variables in the two sets of equations~\eqref{1+3_evol}
and~\eqref{DZ_evol1}.  The details are given in
Appendix~\ref{app:1+3}, where it is shown that:
\begin{subequations} \label{link}
\begin{align}
{\bf D}^2{\mathbb D} &= {}^{(1)}\!D - 3\cH
{}^{(1)}\!\tilde {Q}, \label{link_D}\\
{\bf D}^2{\mathbb Z} &= {}^{(1)}\!Z + ({\bf D}^2 + 3K - \sfrac32{\cal
A}_G){}^{(1)}\!\tilde {Q} \label{link_Z}\\
{\bf D}^2{\bar{\Xi}} &= {}^{(1)}\!{\tilde \Pi}, \label{link_Xi}\\
{\bf D}^2\bar{\Gamma} &= {}^{(1)}\!\tilde{\Gamma} .
\end{align}
\end{subequations}
If we now apply the operator ${\bf D}^2$ to
equations~\eqref{DZ_evol1} and use equations~\eqref{link} then
we obtain precisely equations~\eqref{1+3_evol}, with each
variable by its linear perturbation and each differential
operator replaced by its zeroth order perturbation.

\section{Discussion}
\label{sec:discuss}

In this paper we have presented an efficient way of defining
dimensionless gauge invariants and determining their
inter-relationships, which we have applied to give a unified
account of the various ways of formulating the governing
equations for scalar perturbations of FL cosmologies. In
defining gauge invariants we use our version of Nakamura's
geometrical method, as described in UW (see section 2.1) which
is based on specifying a so-called gauge field $X$, and
normalizing so as to obtain dimensionless quantities.
It turns out that the choice of the spatial part
$X^i$ of the gauge field does not affect
the form of the governing equations for linear perturbations
given in UW and in the present paper.\footnote
{This feature of linear perturbations is due to the fact that the
components of the perturbed Riemann tensor and of the
perturbed stress-energy tensor that are used in deriving
the governing equations are invariant under {\it spatial}
gauge transformations. Indeed, according to Bardeen (1988),
since the background 3-space is homogeneous and isotropic
the perturbations in all physical quantities must be invariant under spatial
gauge transformations. Whether this property holds for
nonlinear perturbations requires further investigation.}
In these papers, however, we find it convenient to fix
the spatial part $X^i$ of the gauge field
according to~\eqref{X_i}, which leads to a  simple form~\eqref{bold_f_split} for the
metric gauge invariant ${\bf f}_{ab}[X]$. This in turn shortens the
calculation of the Riemann gauge invariants in UW (see (B.23)) and of the
gauge-invariant form of the divergence of the stress-energy
tensor in the present paper, equation \eqref{cons_X}.\footnote
{If one does not fix $X^i$ one has to use the general form~\eqref{bold_f_split1}
of ${\bf f}_{ab}[X]$. However,
during the calculations the terms involving
$X^i$ cancel, leading to the same final results.}
 The remaining gauge freedom is then
described by the temporal part $X^0$, which we specify uniquely
by requiring that one of the four basic gauge invariants $
\Psi[X], {\bf B}[X], {\mathbb D}[X]$ and ${\mathbb V}[X]$ be zero.
This approach eliminates the need to express $X^0$ explicitly
in terms of gauge-variant variables, and in particular enables
one to perform a change of gauge without using $X^0$, as in
subsection~\ref{subsection:choice}.
Indeed, although the gauge field $X^a$, which is gauge-variant,
plays an important role in establishing our formalism, we do
not use it in performing calculations, in keeping with our goal
of working exclusively with gauge invariants. Because it
simplifies calculations this approach will facilitate the
extension to second order perturbations.

The coupled system of first order partial differential equations for ${\mathbb D}$ and ${\mathbb
Z}$, given by equations~\eqref{DZ_evol1}, plays a central role
in this paper. To the best of our knowledge they have not been
given in the literature. They arise first of all from the
linearized conservation equations in conjunction with the
linearized Einstein equations, but can also be derived directly
from the latter equations, when they are written in terms of
the uniform curvature gauge invariants as
in~\eqref{scalar_eq_curv}. Further, as shown in
section~\ref{sec:1+3} these equations are essentially
equivalent to the first order governing equations for $D$ and $Z$
that arise in the $1+3$ gauge-invariant approach (see
equations~\eqref{1+3_evol}).

Our derivation of the expressions for the conserved quantities
$\zeta_\rho$ and $\zeta_{\mathrm v}$ deserves comment. We have
shown that the conservation equations for $\zeta_\rho$ and
$\zeta_{\mathrm v}$ are simply two of the first order governing
equations for scalar perturbations when they are written in
terms of the appropriate gauge invariants: the gauge-invariant
expression for $\zeta_\rho$ arises directly from the linearized
conservation equations for the stress-energy tensor, while that
for $\zeta_{\mathrm v}$ arises from the linearized Einstein
equations for the uniform curvature metric gauge invariants.
Other expressions for the conserved quantities are derived
using the method for finding inter-relationships between gauge
invariants given in section~\ref{subsection:choice}. We mention
that $\zeta_{\mathrm v}$ is usually introduced by rewriting the
second order evolution equation~\eqref{bardeen} for the Bardeen
potential $\Psi$ in a first order form, a procedure that
involves a tedious calculation.\footnote{This calculation is
much easier if one uses our factorization
property~\eqref{factorL_s} for the operator $\cL$.} Our use of
the uniform curvature metric gauge invariants avoids the need
for any calculation.

Our discussion of the $1+3$ gauge-invariant approach has
several novel features. An advantage of this approach is that
it is coordinate-free, so that calculations require only
standard operations from differential geometry. This feature
has enabled us to give a particularly concise derivation of the
first order system of governing equations for scalar
perturbations, as given by equations~\eqref{1+3_evol} (see
Appendix \ref{app:1+3}).\footnote{In the usual derivation one
first obtains a system of partial differential equations for the spatial
gradients of the energy density and the Hubble scalar, and then
one takes the spatial divergence to obtain partial differential equations for scalars,
a more lengthy process. See, for example, BDE.} We have also
derived the relation between the variables $({\mathbb D}, {\mathbb
Z})$ in the metric-based approach and the $1+3$ variables $(D, Z)$
(see equation~\eqref {link}). A drawback of the $1+3$ approach
is that the linearization process is conceptually less clear
than in the metric-based approach, relying as it does on "dropping
products of first order terms". In relating the $1+3$ approach
to the metric-based approach it was necessary to regard $D, Z$ and
the differential operators as functions of the perturbation
parameter $\epsilon$ and explicitly calculate their dependence
on $\epsilon$ to linear order. Introducing the perturbation
parameter clarifies the linearization process and points the
way to extending the $1+3$ approach to second order
perturbations.

\subsection*{Acknowledgments}
CU is supported by the Swedish Research Council (VR grant
621-2009-4163). CU also thanks the Department of Applied
Mathematics at the University of Waterloo for kind hospitality.
JW acknowledges financial support from the University of
Waterloo.

\appendix

\section{The Replacement Principle}\label{app:repl}

We define
\be I_a(\epsilon) := {\cal M}^2\, {^{\epsilon} }\bna\!_bT^b\!_a(\epsilon).  \ee
The linear perturbation of $ I_a$, given in equation~\eqref{conservcompi}, can be
written symbolically in the form
\be \label{I}  ^{(1)}I_a = {\mathsf L}_a\left( {\cal M}^2 {^{(1)}}T^b\!_c\,, ^{(1)}\!\!f_{bc}\right),  \ee
where $ {\mathsf L}_a$ is a linear operator.
The replacement principle for the divergence of the stress-energy tensor
states that the gauge invariants associated with $^{(1)}I_a, {^{(1)}}T^a\!_b$
and $ {^{(1)}}f_{ab}$ by $X$-compensation are related by the {\it same} linear operator:
\be \label{I_X}  {\bf I}_a[X] = {\mathsf L}_a\left( {\bf T}^b\!_c[X], {\bf f}_{bc}[X]\right),  \ee
for any gauge field $X$.
If the stress-energy tensor is conserved at zero order ({\it i.e.} $I_a(0)=0$
then $^{(1)}I_a$ is a gauge invariant, and the left sides of~\eqref{I} and~\eqref{I_X} are
equal.

This result is adapted from Nakamura (2005) (see equations (3.90), (3.91) and (3.20)).
 Use of the Replacement Principle in Appendix
\ref{app:derconserved} makes the transition from gauge-variant to
gauge-invariant equations particularly easy and transparent.

\section{Derivation of the conservation
equations}\label{app:derconserved}

In this Appendix we give the derivation of the linearized
conservation equations in the form~\eqref{cons_X}, using the
methods developed in UW (see in particular Section 2 and
Appendix B). We express the covariant derivative
${}^\epsilon\bna\!_a$ of the metric $g_{ab}(\epsilon)$  in
terms of the covariant derivative ${}^0\!\bar{\bna}\!_a$ of the
conformal background metric $\gamma_{ab}$ as follows:
\be \label{def_Q} {}^\epsilon\bna\!_a A^{b}\!_{c}(\epsilon) =
{}^0\!\bar{\bna}\!_a A^{b}\!_{c}(\epsilon) +
Q^{b}\!_{ad}(\epsilon)A^{d}\!_{c}(\epsilon) -
Q^{d}\!_{ac}(\epsilon)A^{b}\!_{d}(\epsilon). \ee
The object $Q^a\!_{bc}(\epsilon)$ is written as the sum of
two parts:
\be \label{Qcosmo}  Q^a\!_{bc}(\epsilon) =
\bar{Q}^a\!_{bc}(\epsilon) + \tilde{Q}^a\!_{bc}(\epsilon), \ee
where
\begin{subequations}
\begin{align}
\bar{Q}^a\!_{bc} (\epsilon) &:= 2\delta^a\!_{(b} r_{c)} - \bar{g}^{ad}(\epsilon)\bar{g}_{bc}(\epsilon) r_d, \qquad
\text{with} \qquad  r_a: = {}^0\!\bar{\bna}\!_a (\ln a), \label{Qbar}\\
\tilde{Q}^a\!_{bc}(\epsilon) &:= \sfrac{1}{2}\,\bar{g}^{ad}(\epsilon)\left({}^0\!\bar{\bna}\!_{c}\,\bar{g}_{db}(\epsilon)
- {}^0\!\bar{\bna}\!_{d}\,\bar{g}_{bc}(\epsilon) +
{}^0\!\bar{\bna}\!_{b}\,\bar{g}_{cd}(\epsilon)\right). \label{Qtilde}
\end{align}
\end{subequations}
It follows from~\eqref{Qbar} and~\eqref{Qtilde},
in conjunction with ${}^0\!\bar{\bna}\!_a \gamma_{bc}=0$, that at
zeroth and first order we obtain
\begin{subequations} \label{Q}
\begin{xalignat}{2}
{}^{(0)}\!\bar{Q}^a\!_{bc} &= 2\delta^a\!_{(b} r_{c)} -
\gamma^{ad}\gamma_{bc}r_d, &\quad
{}^{(0)}\!\tilde{Q}^a\!_{bc} &= 0,\label{Q0}\\
{}^{(1)}\!\bar{Q}^a\!_{bc} &= (f^{ad}\gamma_{bc} -
\gamma^{ad}f_{bc})r_d ,&\quad {}^{(1)}\!\tilde{Q}^a\!_{bc} &=
\sfrac{1}{2}\,\gamma^{ad}\left({}^0\!\bar{\bna}\!_{c}\,f_{db} -
{}^0\!\bar{\bna}\!_{d}\,f_{bc} +
{}^0\!\bar{\bna}\!_{b}\,f_{cd}\right).\label{Q1}
\end{xalignat}
\end{subequations}

Consider tensors $A^a\!_b(\epsilon)$ such that $\lambda^2
A^a\!_b(\epsilon)$ is dimensionless, where $\lambda>0$ is a
background quantity with dimension $length$. As follows
from~\eqref{def_Q}, the equation $0 =
\lambda^{2}\,\,{}\!^\epsilon\bna_b A^b\!_a(\epsilon)$ can be
written as
\begin{equation}
0 = \left({}^0\!\bar{\bna}\!_b - 2s_b\right)\lambda^2 A^b\!_a(\epsilon)
+ 2Q^c\!_{b[c}(\epsilon)\lambda^2 A^b\!_{a]}(\epsilon),
\end{equation}
where
\be  s_a: = {}^0\!\bar{\bna}\!_a (\ln \lambda), \ee
which yields the following zeroth and first order expressions
\begin{subequations}\label{conserv}
\begin{align}
0 &= \left({}^0\!\bar{\bna}\!_b - 2s_b\right)\lambda^2 {}^{(0)}\!A^b\!_a
+ 2{}^{(0)}\!\bar{Q}^c\!_{b[c}\lambda^2 {}^{(0)}\!A^b\!_{a]},\label{conserva}\\
0 &= \left({}^0\!\bar{\bna}\!_b - 2s_b\right)\lambda^2 {}^{(1)}\!A^b\!_a
+ 2{}^{(0)}\!\bar{Q}^c\!_{b[c}\lambda^2 {}^{(1)}\!A^b\!_{a]} +
2{}^{(1)}\!Q^c\!_{b[c}\lambda^2 {}^{(0)}\!A^b\!_{a]}.\label{conservb}
\end{align}
\end{subequations}

We now specialize $A^a\!_b(\epsilon)$ to a stress-energy tensor
$T^a\!_b(\epsilon)$ that obeys the background symmetries, {\it
i.e.} that satisfies~\eqref{T_0}, and  choose the normalizing
factor $\lambda$ as in equation~\eqref{c_M}, {\it i.e.}
$\lambda = {\cal M}$. We also assume that $T^a\!_b(\epsilon)$
satisfies a conservation law of the form ${}^\epsilon\bna\!_b
T^b\!_a(\epsilon) = 0$. Relative to local coordinates we obtain
\begin{equation}
r_\alpha =\cH\delta^0\!_\alpha,  \qquad
s_\alpha = \sfrac32{\cal H}(1+\cC_T^2)\delta^0\!_\alpha.
\end{equation}
On substituting from~\eqref{Q} the zeroth order
expression~\eqref{conserva} yields
equations~\eqref{conserved0}, and the temporal and spatial
components of the first order expression~\eqref{conservb}
assume the following form:
\begin{subequations}\label{conservcompi}
\begin{align}
0 &= \partial_\eta({\cal M}^2\,{}^{(1)}\!T^0\!_0 - \sfrac12 f^i\!_i)
+ {\bf D}_i ({\cal M}^2\,{}^{(1)}\!T^i\!_0) -
{\cal H}{\cal M}^2\,\left({}^{(1)}\!T^i\!_i +
3 \cC_T^2 {}^{(1)}\!T^0\!_0 \right),\\
0 &= (\partial_\eta - 3{\cal H}\cC_T^2)({\cal M}^2\,{}^{(1)}\!T^0\!_i) +
{\bf D}_j({\cal M}^2\,{}^{(1)}\!T^j\!_i)  -
{\cal H}\gamma_{ij}\,{\cal M}^2\,{}^{(1)}\!T^j\!_0
- \sfrac12{\bf D}_i f_{00} + {\cal H}f_{0 i} .
\end{align}
\end{subequations}
We simplify these equations by first expressing
${}^{(1)}\!{T}^i\!_0$ in terms of ${}^{(1)}\!{T}^0\!_i$:
\be {\cal M}^2{}^{(1)}\!{T}^i\!_0 =
-\gamma^{ij}\left(\,{\cal M}^2{}^{(1)}\!{T}^0\!_j - f_{0j}\right),
\ee
and decomposing ${}^{(1)}\!T^i\!_j$ into its tracefree part and its
trace using~\eqref{GT_hat}. We then introduce the
intrinsic gauge invariants ${\hat {\mathbb T}}^j\!_i ,
{\mathbb T}_i $ and $ {\mathbb T}$ as defined by~\eqref{GT},
expressing the trace ${}^{(1)}\!T^i\!_i$ in terms of ${\mathbb T}$.
As a result of these changes equations~\eqref{conservcompi} yield:
\begin{subequations}
\begin{align}
0 &= \partial_\eta({\cal M}^2\,{}^{(1)}\!T^0\!_0 - \sfrac12 f^i\!_i)
- {\bf D}^i({\cal M}^2\,{}^{(1)}\!T^0\!_i - f_{0i}) - 3{\cal H}{\mathbb T},\label{cons_1}\\
0 &= (\partial_\eta + {\cal H})({\cal M}^2\,{}^{(1)}\!T^0\!_i)  -
\sfrac12{\bf D}_i f_{00} + {\cC}_T^2 {\mathbb T}_i  +
{\bf D}_j{\hat {\mathbb T}}^j\!_i + {\bf D}_i {\mathbb T}.\label{cons_2}
\end{align}
\end{subequations}
We now apply the Replacement Principle to these equations,
which entails performing the following replacements:
\begin{subequations}\label{replace}
\be  f_{00} \rightarrow {\bf f}_{00}[X], \qquad f_{0i}
\rightarrow {\bf f}_{0i}[X], \qquad f_{ij} \rightarrow {\bf
f}_{ij}[X], \ee
\be {\cal M}^2\,{}^{(1)}\!T^0\!_0 \rightarrow {\mathbb
T}^0\!_0[X], \qquad {\cal M}^2\,{}^{(1)}\!T^0\!_i \rightarrow
{\mathbb T}^0\!_i[X].  \ee
\end{subequations}
On substituting from~\eqref{bold_f_split} and~\eqref{T_i} and
noting that
\be  \label{div_1} {\bf D}^i  {\bf f}_{0i}[X] = {\bf D}^2 {\bf
B}[X],  \qquad {\bf D}^i {\mathbb T}^0\!_i [X] = {\bf D}^2
{\mathbb V}[X],  \ee
equation~\eqref{cons_1} assumes the form~\eqref{cons0}. After
performing the replacements~\eqref{replace} in~\eqref{cons_2}
we apply the operator ${\bf D}^i$ in order to extract the
scalar mode. We then substitute from~\eqref{bold_f_split}
and~\eqref{T_i}, noting~\eqref{div_1} and the fact
that\footnote{The third equality follows from the identity
(B.39e) in UW.}
\be {\bf D}^i {\mathbb T}_i = {\bf D}^2 {\mathbb D},  \qquad {\bf
D}^i{\bf D}_j{\hat {\mathbb T}}^j\!_i = {\bf D}^i{\bf D}_j {\bf
D}^j\!_i {\bar \Pi} = {\bf D}^2 {\bar \Xi},  \ee
where ${\bar \Xi}$ is defined by~\eqref{Xi}.
The result is that~\eqref{cons_2} assumes the form ${\bf D}^2
{\mathbb C} = 0$. Since we are assuming,  as in UW,  that the
inverse operator of ${\bf D}^2 $ exists, we obtain ${\mathbb C} =
0$, which is precisely the desired equation~\eqref{consi}.

\section{Derivation of the $1+3$ perturbation equations}
\label{app:1+3}

\subsection{Derivation of the evolution equations}
\label{app:1+3deriv}

\subsubsection*{The evolution equation for $D$}

We begin with the conservation equations for the stress-energy
tensor~\eqref{T_ab}, linearized by dropping products of first
order quantities\footnote{See, for example, Wainwright and
Ellis (1997), equations (1.48) and (1.49), after multiplying by
$a$ to change the dot derivative to prime.}:
\begin{subequations}
\begin{align}
\rho' &= -3aH(\rho + p) - {\tilde {\bf D}}^a q_a, \label{rho'}\\
h_a\!^b q_b' &= - 4aH q_a - {\tilde {\bf D}}_a p -
(\rho + p)a{\dot u}_a - {\tilde {\bf D}}_b \pi^b\!_a. \label{q'}
\end{align}
\end{subequations}
In these equations the differential operators $'$ and
${\tilde {\bf D}}_a$ are defined by~\eqref{operators1} and \eqref{operators4}.
We require the zero order version of~\eqref{rho'}
which we write in the form\footnote{For a background scalar
${}^{(0)}\!A' $ is the ordinary derivative with respect to conformal
time $\eta$.}
\be {\cal M}^2 \,{}^{(0)}\!\rho' = -3{\cal H},
\label{rho_prime} \ee
which leads to the evolution equation for $ {\cal M}^2$:
\be  ({\cal M}^2)'  = 3(1 + {\cal C}_T^2) {\cal H}  {\cal M}^2. \ee

We apply the operator $ {\cal M}^2{\tilde {\bf D}}^2$
to~\eqref{rho'} and the operator $ {\cal M}^2{\tilde {\bf D}}^a$
to~\eqref{q'} and then linearize, obtaining
\begin{subequations}
\begin{align}
D' - 3\cH {\cal C}_T^2 D + Z &= - 3\cH\!\left({\tilde {\bf D}}^a(a{\dot u}_a) +
P\right) - {\tilde {\bf D}}^2 {\tilde Q},\label{int1}\\
{\tilde {\bf D}}^a(a{\dot u}_a) + P &=
-( {\tilde \Upsilon} + {\tilde \Pi}).\label{int2}
\end{align}
\end{subequations}
On substituting~\eqref{int2} in~\eqref{int1} we obtain
the evolution equation~\eqref{D_evol} for $D$.

In deriving~\eqref{int1} and~\eqref{int2} we use the following
linearized commutativity properties:
\begin{subequations}
\be  \label{bna^2_prime}  {\tilde {\bf D}}^2 (A') =
\left({\tilde {\bf D}}^2 A\right)\!{'} -
{}^{(0)}\!A'\!\left({\tilde {\bf D}}^a(a{\dot u}_a)\right), \ee
where $A$ is any scalar field, and
\be  {\tilde {\bf D}}_a A_b' =  ({\tilde {\bf D}}_a A_b)',    \ee
\end{subequations}
where $A_a$ is any covariant vector field.
In differentiating products of perturbed quantities such as
$\rho H$, $Hq_a$ and $(\rho + p){\dot u}_a$ we use the
following expansion to linear order:
\be \label{AB_lin}  AB = {}^{(0)}\!AB + {}^{(0)}\!BA -  {}^{(0)}\!A {}^{(0)}\!B,  \ee
where $A$ and $B$ are geometric quantities with background
values ${}^{(0)}\!A $ and ${}^{(0)}\!B$, one of which may be
zero.

\subsubsection*{The evolution equation for $Z$}

We begin with the linearized Raychaudhuri equation written
in the form
\be  \label{Raychaud}  3(aH' + a^2H^2) - {\tilde {\bf D}}^a(a{\dot u}_a) +
 \sfrac12 a^2 (\rho + 3p) = 0. \ee
We use~\eqref{AB_lin} with $A=B=H$ to write $a^2H^2 = 2{\cal
H}(aH) - {\cal H}^2$, where ${\cal H} :=a{}^{(0)}\!H, $ and
use~\eqref{int2} to eliminate ${\dot u}_a$. We then  apply the
operator ${\tilde {\bf D}}^2$ to~\eqref{Raychaud}. After
using~\eqref{bna^2_prime} and the definitions of $D,Z$ and $P$
we obtain\footnote{In doing calculations such as these one
should keep in mind that ${}^{(3)}{\bna}\!_a(a')\neq 0$, where
$'$ is defined by~\eqref{operators1},  even though we have
chosen $a$ to be the background scale factor ({\it i.e.}
${}^{(3)}{\bna}\!_a(a) = 0$.}
\be \label{int3} Z' + {\cal H}Z + \sfrac12{\cal A} D =  -
({\tilde {\bf D}}^2 + 3K)P -
({\tilde {\bf D}}^2 + 3K - \sfrac32{\cal A}) ({\tilde \Upsilon} + {\tilde \Pi}).  \ee
In deriving this equation we have also used~\eqref{cal_C} and~\eqref{Hprime}.
We finally use~\eqref{tildeGamma} to express $P$ in~\eqref{int3} in terms
of $D$ and ${\tilde \Gamma}$, which gives the evolution equation
\eqref{Z_evol} for $Z$.

\subsection{Relation between the $1+3$ and the metric-based approaches}

\subsubsection*{Fundamental 4-velocity and energy flow vector}
\label{app:1+3metric}

We begin with the decomposition of the stress-energy tensor
with respect to a unit timelike vector field $u^a$, which is
given by~\eqref{T_ab}. The Stewart-Walker lemma implies that
the linear perturbation ${}^{(1)}\!{q}_a$ is a gauge invariant.
Since ${}^{(0)}\!u^a = a^{-1} \delta^a\!_0$ and ${}^{(0)}\!u_a
= - a\, \delta^0\!_a$, it follows that ${}^{(1)}\!{q}_0 = 0$,
and hence that
\begin{subequations}  \label{TvQ}
\be {}^{(1)}\!T^0\!_0 = - {}^{(1)}\!\rho, \quad
{\cal M}^2\, {}^{(1)}\!T^0\!_i = v_i + {\bar{\mathbb Q}}_i, \ee
where
\be a v_i:={}^{(1)}\!{u}_i,  \quad a\bar{\mathbb Q}_i :=  {\cal
M}^2\,{}^{(1)}\!{q}_i.    \ee
\end{subequations}
It follows from~\eqref{TvQ} and~\eqref{TiX} that
\begin{subequations}
\be  \label{TvQ_gi}  {\mathbb T}^0\!_i[X] =  {\bf v}_i[X] +
\bar{\mathbb Q}_i,  \ee
where
\be \label{u_gi} {\bf v}_i[X] = v_i + {\bf D}_i X^0.  \ee
\end{subequations}
We decompose ${\bar{\mathbb Q}}_i, {\bf v}_i$  and $v_i$ according to
\begin{subequations}  \label{vQ_decomp}
\be
 {\bar{\mathbb Q}}_i = {\bf D}_i {\bar{\mathbb Q}}  + \tilde{{\mathbb Q}}_i,
\qquad {\bf v}_i[X] = {\bf D}_i {\bf v}[X] + {\tilde {\bf
v}}_i, \qquad v_i= D_i v +{\tilde v}_i,   \ee
with
\be {\bf D}^i \tilde{{\mathbb Q}}_i = 0,  \qquad  {\bf D}^i
\tilde{{\bf v}}_i = 0,  \qquad  {\bf D}^i \tilde{ v}_i = 0.
\ee
\end{subequations}
It now follows from~\eqref{hybrid_T},~\eqref{TvQ_gi}
and~\eqref{vQ_decomp} that
\begin{equation}
{\mathbb V}[X] = {\bf v}[X]  +{\bar{\mathbb Q}},  \qquad
{\mathbb V}_i ={\tilde {\bf v}}_i +\tilde{{\mathbb Q}}_i . \label{pf_V}
\end{equation}
Thus if the preferred timelike vector field $u^a$ is an
eigenvector of the stress-energy tensor, {\it i.e.} if the
energy transfer vector $q^a$ is zero, then the stress-energy
gauge invariants ${\mathbb V}[X]$ and ${\mathbb V}_i$ equal the gauge
invariants $ {\bf v}[X]$ and $\tilde{\bf v}_i$ associated with
$u^a$. In addition it follows from~\eqref{u_gi} and~\eqref{vQ_decomp}
that
\be \label{gi_v} {\bf v}[X] = v + X^0,  \qquad  {\tilde {\bf
v}}_i = {\tilde v}_i.  \ee

\subsubsection*{Spatial gradient and Laplacian of a scalar}

We have seen that the $1+3$ approach to
cosmological perturbations is based on the spatial gradient and
Laplacian of the density $\rho$ and the Hubble scalar $H$. We now
define these quantities for a scalar field of given
dimension, using a background normalization factor $\lambda$ of
dimension \emph{length}. Let $f$ be a scalar such that
$\lambda^n f$ is dimensionless, and whose unperturbed value is
a function only of $\eta$. We define the dimensionless spatial
gradient and spatial Laplacian of $f$ according
to
\be \label{grad,lapl} F_a := \frac{1}{a} {\tilde {\bf
D}}_a(\lambda^n f), \qquad F :=  {\tilde {\bf D}}^2(\lambda^n
f),  \ee
using the notation~\eqref{laplacian}.

Our goal is to relate the linear perturbation of $F_a$ and $F$
to the linear perturbation of $f$. Regarding all perturbed
quantities as functions of the perturbation parameter
$\epsilon$, we write
\be \label{epsilon_dep} F_a(\epsilon) :=  h_a\!^b(\epsilon)
{}^{\epsilon} {\bna}\!_a(\lambda^n f(\epsilon) ), \qquad
F(\epsilon) :=  a^2 g^{ab}(\epsilon) h_a\!^c(\epsilon)
{}^{\epsilon}{\bna}\!_c F_b(\epsilon), \ee
with $f(\epsilon) =  {}^{(0)}\!f + \epsilon\,{}^{(1)}\!f +
\dots\,$, {\it etc}. A straightforward calculation
yields\footnote{We note that $^{\epsilon}{\bna}\!_a A(\epsilon)
= {}^{0}{\bar {\bna}\!_a}A(\epsilon)$ for a scalar A, and that
${}^0\bar{\bna}\!_i A = {\bf D}_i A $ and ${}^0\bar{\bna}\!_0 A
=\partial_\eta A.$}
\begin{subequations}\label{step1}
\begin{xalignat}{3}
{}^{(0)}\!F_a &= 0,  &\, {}^{(1)}\!F_0 &= 0,  &\, {}^{(1)}\!F_i
&= \lambda^n\!\left({\bf D}_i {}^{(1)}\!f + {}^{(0)}\!f'
{v}_i\right)\!,\\
{}^{(0)}\!F &= 0,  &\,  {}^{(1)}\!F &={\bf D}^i\,
{}^{(1)}\!F_i. &&
\end{xalignat}
\end{subequations}
We note that the background values ${}^{(0)}\!F_a$ and
${}^{(0)}\!F$ are zero due to our assumption that
${}^{(0)}\!f = {}^{(0)}\!f(\eta)$. On account of the
Stewart-Walker lemma the linear perturbations ${}^{(1)}\!F_i $ and
${}^{(1)}\!F$ are gauge-invariant. We can write them in a
manifestly gauge-invariant form by noting that
\be  \label{step2} \lambda^n\left({\bf D}_i{^{(1)}\!f } + ^{(0)}\!\!f' v_i\right) =
{\bf D}_i{\bf f}[X] +  \lambda^n\,{^{(0)}\!f' }{\bf v}_i[X],     \ee
where ${\bf f}[X] $ is the gauge invariant associated with $f$
by $X$-compensation and ${\bf v}_i[X]$ is given
by~\eqref{u_gi}. It follows from~\eqref{step1},~\eqref{step2}
and~\eqref{vQ_decomp} that
\be  \label{laplacian1}   {}^{(1)}\!F = {\bf D}^2\!\left( {\bf
f}[X] + \lambda^n\,{^{(0)}\!f' }\,{\bf v}[X]\right).  \ee
For future use we choose $X=X_{\mathrm v}$ in~\eqref{laplacian1}
and use the fact that ${\bf v}[X_{\mathrm v}] = - {\bar{\mathbb Q}}$, as follows
from~\eqref{pf_V}. Equation~\eqref{laplacian1} assumes the form
\be  \label{laplacian2}  {}^{(1)}\!F = {\bf D}^2\!\left( {\bf
f}[X_{\mathrm v}] - \lambda^n\,{}^{(0)}\!f'\,\bar{\mathbb
Q}\right).  \ee

\subsubsection*{Relation between the variables}

We need an expression for the gauge-invariant linear perturbation
${\bf H}[X]$ of the Hubble scalar $H$ of the preferred congruence,
which is defined by
\be {\bf H}[X] = a(^{(1)}H - ^{(0)}\!H' X^0),  \ee
in accordance with the general definition~\eqref{bold_A}.
It follows
from the expression (B.41a) for $a^{(1)}H$ in UW, in
conjunction with~\eqref{metric_gi} and~\eqref{gi_v},
that\footnote
 {This is another example of a Replacement Principle, where
 an equation remains valid when each gauge-variant quantity is
 replaced by the associated gauge invariant defined using
 $X$-compensation.}
\be\label{H_X}  {\bf H}[X] = \sfrac13 {\bf D}^2({\bf v}[X] -
{\bf B}[X]) - (\partial_\eta \Psi[X] + \cH \Phi[X]).  \ee
We have also used the fact that
\be  \label{Hprime}  a{}^{(0)}\!H' = {\cal H}' - {\cal H}^2
= - (\sfrac12 \cA_G - K),  \ee
the second equality following from~\eqref{A_G}.

First we note that
\be\label{H1}  {\bf v}[X] - {\bf B}[X] = {\mathbb V}[X] - {\bf
B}[X] - {\bar{\mathbb Q}} = {\mathbb V} - {\bar {\mathbb Q}},  \ee
as follows from~\eqref{pf_V} and the $X$-independent gauge
invariant $[{\mathbb V},{\bf B}]$ in \eqref{indep_X}. Second we can use the
transition rules~\eqref{transition} and~\eqref{transition2} to
show that the gauge invariant
\be\label{PsiPhiV} \partial_\eta \Psi[X] + \cH \Phi[X] + (-\cH'
+ \cH^2){\mathbb V}[X],  \ee
is $X$-independent. Evaluating this expression for
$X^0 = X^0_{\mathrm v}$ and $X^0 = X^0_{\mathrm p}$ yields
\be\label{H2} \partial_\eta \Psi_{\mathrm v} + \cH
\Phi_{\mathrm v} = -K{\mathbb V},   \ee
where we have used the linearized Einstein
equation~\eqref{V_P}, the background Einstein equation $\cA_T =
\cA_G$ and the definition~\eqref{A_G} of $\cA_G.$ Finally we
choose $X^0 = X^0_{\mathrm v}$ in~\eqref{H_X} and substitute from~\eqref{H1}
and~\eqref{H2} to obtain the desired result that
\be  \label{H_V}  3{\bf H}_{\mathrm v} = ( {\bf D}^2 + 3K){\mathbb
V} - {\bf D}^2 {\bar{\mathbb Q}} = {\mathbb Z} -  {\bf D}^2
{\bar{\mathbb Q}},  \ee
the second equality following from~\eqref{Bbb_Z}.

We can now use~\eqref{laplacian2} to relate the perturbations
of the variables $D$ and $Z$ in the  $1+3$ approach to the
corresponding variables ${\mathbb D}$ and ${\mathbb Z}$ in the
metric-based approach. First choose $f=\rho$, $\lambda = {\cal M}$
and $n=2$ in~\eqref{laplacian2}, and use~\eqref{TvQ},
\eqref{rho_prime}, \eqref{DX} and~\eqref{DV} to obtain
\be \label{D1_int} {}^{(1)}\!D = {\bf D}^2({\mathbb D} +3{\cal H}{\bar{\mathbb Q}}).  \ee
Second, choose  $f= H$, $\lambda = a$ and $n=1$
in~\eqref{laplacian2}. Equations~\eqref{H_V} and~\eqref{Hprime}
then lead to\footnote{Note the factor 3 in the definition of
$Z$ in~\eqref{DZ_defn} compared to~\eqref{grad,lapl}, so that
$Z=3F$ and ${}^{(1)}\!Z=3{}^{(1)}\!F$ .}
\be \label{Z1_int}  {}^{(1)}\!Z = {\bf D}^2\!\left({\mathbb Z} - (
{\bf D}^2 + 3K - \sfrac32{\cal A}_G){\bar{\mathbb Q}}\right). \ee
We also need the relation
\be \label{Q1_int}  {}^{(1)}\!{\tilde Q} = {\bf D}^2 {\bar{\mathbb Q}},  \ee
which can be derived from the definition~\eqref{tildeQ} of
${\tilde Q}$ and the definition~\eqref{vQ_decomp} of
${\bar{\mathbb Q}}$. The desired equations~\eqref{link_D}
and~\eqref{link_Z} now follow immediately from~\eqref{D1_int}
and~\eqref{Z1_int} in conjunction with~\eqref{Q1_int}. Further,
equations~\eqref{link_Xi} can be derived from the
definitions~\eqref{tildeGamma} and~\eqref{tildeQ} of ${\tilde \Gamma}$
and ${\tilde \Pi}$. Finally, equation~\eqref{operators3} can be
derived in a similar manner using~\eqref{operators2} and the
footnote following~\eqref{epsilon_dep}.

\section*{References}

\noindent Bardeen, J. M. (1980) Gauge-invariant cosmological
perturbations, {\it Phys. Rev. D} {\bf 22}, 1882-1905.\\

\noindent Bardeen, J. M. (1988) Cosmological
perturbations, from quantum fluctuations to large scale structure,
in {\it Cosmology and Particle Physics}, edited by Li-Zhi Fang and A. Zee,
pages 1-64 (Gordon and Breach, New York).  \\

\noindent  Bardeen, J.M., Steinhardt, P.J. and Turner, M.S. (1983)
Spontaneous creation of almost scale-free density
perturbations in an inflationary universe, {\it Phys. Rev. D} {\bf 28}, 679-693.  \\

\noindent Brandenberger, R. and Khan, R. (1984) Cosmological perturbations
in inflationary-universe models,  {\it Phys. Rev. D} {\bf 29}, 2172-2190.\\

\noindent Bruni, M., Dunsby, P.K.S. and Ellis, G.F.R. (1992a)
Cosmological perturbations and the meaning of gauge-invariant
variables, {\it Astrophysical J.} {\bf 395}, 34-53.\\

\noindent Bruni, M., Dunsby, P.K.S. and Ellis, G.F.R. (1992b)
Gauge-invariant perturbations in a scalar field dominated
universe,  {\it Class. Quantum Grav.} {\bf 9}, 921-945.\\

\noindent Bruni, M., Matarrese, S., Mollerach, S., and Sonego,
S. (1997) Perturbations of spacetime: gauge transformations and
gauge-invariance at second order and beyond,
{\it Class. Quantum Grav.} {\bf 14}, 2585-2606.\\

\noindent  Dunsby, P.K.S., Bruni, M., and Ellis, G.F.R. (1992)
Covariant perturbations in a multifluid cosmological
medium, {\it Astrophysical J.} {\bf 395}, 54-74.\\

\noindent Durrer, R. (2008) {\it The Cosmic Microwave Background},
Cambridge University Press. \\

\noindent Ellis, G.F.R. and Bruni, M. (1989), Covariant and
gauge-invariant approach to cosmological density
fluctuations, {\it Phys. Rev. D} {\bf 40}, 1804-1818.\\

\noindent  Ellis, G.F.R., Bruni, M. and Hwang, J (1990)
Density-Gradient-vorticity relation in perfect-fluid Robertson-Walker
perturbations,  {\it Phys. Rev. D} {\bf 42}, 1035-1046.\\

\noindent Ellis, G.F.R., Hwang, J. and Bruni, M. (1989)
Covariant and gauge-independent perfect fluid Robertson-Walker
perturbations, {\it Phys. Rev. D} {\bf 40}, 1819-1826.\\

\noindent Hawking, S.W. (1966) Perturbations of an expanding
universe, {\it Astrophysical J.} {\bf 145}, 544-54.\\

\noindent Hwang, J. (1991) Perturbations of the Robertson-Walker
space: multicomponent sources and generalized gravity,
 {\it Astrophysical J.} {\bf 375}, 443-462.\\

\noindent Hwang, J. and Noh, H. (1999) Relativistic
Hydrodynamic Cosmological Perturbations,
{\it General Relativity and Gravitation} {\bf 31},1131-1146.  \\

\noindent Kodama, H. and Sasaki, M. (1984) Cosmological
Perturbation Theory, {\it Prog. Theoret. Phys. Suppl. } {\bf 78},1-166.\\

\noindent
Langlois, D. and Vernizzi, F. (2005) Conserved nonlinear quantities
in cosmology, {\it Phys. Rev. D} {\bf72}, 103501 (1-9).  \\

\noindent Malik, K. A. and Wands, D. (2009)
Cosmological perturbations, {\it Physics Reports} {\bf 475}, 1-51.\\

\noindent Mukhanov, V. (2005) {\it Physical Foundations of Cosmology},
Cambridge University Press. \\

\noindent Mukhanov, V. F., Feldman, H. A. and Brandenberger,
R. H. (1992) Theory of cosmological perturbations, {\it Physics Reports} {\bf 215}, 203-333.\\

\noindent Nakamura, K. (2003) Gauge Invariant Variables in
Two-Parameter Nonlinear Perturbations, {\it Prog. Theor. Phys.}
{\bf 110}, 723-755. \\

\noindent Nakamura, K. (2005) Second Order Gauge Invariant
Perturbation Theory, {\it Prog. Theor. Phys.} {\bf 113},
481-511. \\

\noindent Nakamura, K. (2007) Second Order Gauge Invariant
Cosmological Perturbation Theory, {\it Prog. Theor. Phys.} {\bf 117},
17-74. \\

\noindent Tsagas C. G., Challinor A. and Maartens R. (2008)
Relativistic cosmology and large-scale structure, {\it Physics
Reports} {\bf 465}, 61-147. \\

\noindent Uggla, C. and Wainwright, J. (2011) Cosmological
 Perturbation Theory Revisited, {\it Class. Quantum Grav.} {\bf 28},
 175017(26pp).  \\

 \noindent Wainwright, J. and Ellis, G.F.R. (1997) {\it Dynamical Systems in
 Cosmology}, Cambridge University Press.  \\

\noindent
Wands, D. Malik, K. A. Lyth, D. H. and Liddle, A. R. (2000),
New approach to the evolution of cosmological perturbations on
large scales, {\it Phys. Rev. D} {\bf 62}, 043527 (1-8).       \\

\noindent Weinberg, S. (2008) {\it Cosmology}, Oxford University Press. \\

\noindent
Woszczyna, A. and Kulak, A. (1989) Cosmological
perturbations - extension of Olson's gauge-invariant method,
{\it Class. Quantum Grav.} {\bf 6}, 1665-1671.  \\

\end{document}